# Quasicrystal bulk and surface energies from density functional theory

Woohyeon Baek[1], Sambit Das[2], Shibo Tan[1], Vikram Gavini[1,2], Wenhao Sun[1*]

[1]Department of Materials Science and Engineering, University of Michigan; Ann Arbor, MI, USA
[2]Department of Mechanical Engineering, University of Michigan; Ann Arbor, MI, USA

*Corresponding Author: whsun@umich.edu

**Abstract:** Are quasicrystals stable or metastable? Density functional theory (DFT) is often used to evaluate thermodynamic stability, but quasicrystals are long-range aperiodic and their energies cannot be calculated using conventional *ab initio* methods. Here, we perform first-principles calculations on quasicrystal nanoparticles of increasing sizes, from which we can directly extrapolate their bulk and surface energies. Using this technique, we determine with high confidence that the icosahedral quasicrystals $ScZn_{7.33}$ and $YbCd_{5.7}$ are ground-state phases—revealing that translational symmetry is not a necessary condition for the $T = 0$ K stability of inorganic solids. Although we find the $ScZn_{7.33}$ quasicrystal to be thermodynamically stable, we show on a mixed thermodynamic and kinetic phase diagram that its solidification from the melt is nucleation-limited, which illustrates why even stable materials may be kinetically challenging to grow. Our techniques here broadly open the door to first-principles investigations into the *structure-bonding-stability* relationships of aperiodic materials.

Quasicrystals are a mesmerizing and provocative class of materials. With their long-range aperiodicity and forbidden rotational symmetries, the discovery of quasicrystals[1] forced solid-state chemists to reconsider long-standing assumptions regarding crystallinity, bonding, and materials formation.[2,3] Although over 100 intermetallic quasicrystals have been experimentally characterized,[4,5] the underlying mechanisms driving quasicrystal formation are still not fully understood. One longstanding and fundamental question is: *Are quasicrystals enthalpy stabilized or entropy stabilized*?[3] In other words, are quasicrystals $T$ = 0 K thermodynamic ground states; or are they enthalpically metastable[6] but become entropy-stabilized at high temperatures by phasons, phonons, aperiodic substructural tilings, or other forms of configurational disorder?[7–10]

Prior assertions on the thermodynamic stability or metastability of quasicrystals have relied primarily on indirect arguments. For example, the binary $YbCd_{5.7}$ and $ScZn_{7.33}$ quasicrystals are argued to be stable due to their persistence under thermal annealing at elevated temperatures[11,12]—however this thermal stability could be kinetic rather than thermodynamic stability. Various quasicrystals are also claimed to be metastable, due to their structural disorder from collective atomic fluctuations known as 'phasons',[13,14] Hume-Rothery derived arguments about their electron/atom concentrations,[15,16] or their observation as transient intermediates during solidification.[17–19] However, these arguments are also indirect measures of thermodynamic stability.

Density functional theory (DFT) calculations are routinely used to evaluate the thermodynamic stability of inorganic solids from first-principles. However, DFT calculations rely on periodic boundary conditions, meaning that conventional DFT calculations cannot be used to calculate quasicrystals. DFT has been applied to quasicrystal approximants,[20,21] which are crystalline phases with the same substructural building-blocks of quasicrystals. However, these approximants do not exhibit the essential quasicrystalline characteristic of aperiodicity. Molecular dynamics simulations with isotropic multi-well interatomic potentials have been able to produce quasicrystalline phases *in silico*,[22–26] which is remarkable given the simplicity of these pair potentials. However, it is unclear if these isotropic interatomic potentials capture the true quantum-chemical bonding and steric interactions in real intermetallic quasicrystals.

Here, we present a first-principles method to directly evaluate the bulk and surface energies of quasicrystals from density functional theory. Our technique leverages the fact that the volumes of nanoparticles scale with the number of atoms, $N$, but their surface areas scale with $N^{2/3}$. The total energy of a nanoparticle can be written as a sum of the bulk energy and the surface energy: $E_{NP} = E_{bulk} + \gamma A$, where $\gamma$ is the surface energy and $A$ is the surface area. Normalizing by the number of atoms, $N$, gives:

$$\frac{E_{NP}}{N} = \frac{E_{bulk}}{N} + \gamma\eta\rho^{-\frac{2}{3}} \cdot N^{-\frac{1}{3}}$$

where $\rho$ is the atomic density, and $\eta$ is the dimensionless shape factor. By calculating the energies of quasicrystal nanoparticles of increasing sizes, we can fit a linear regression to this relation and directly extract both the bulk and surface energies of a quasicrystal. This size-dependent energy scaling relationship has previously been used to determine surface energies, either from DFT surface slab calculations[27–29] or melt-drop solution calorimetry experiments.[30,31] Here, we repurposed this relation to extract the bulk energy of an aperiodic material from first-principles calculations.

## Bulk stability of icosahedral quasicrystals

Using this technique, we investigated the thermodynamic stability of the Tsai-type binary icosahedral quasicrystals (iQC) $ScZn_{7.33}$ and $YbCd_{5.7}$, whose atomistic structures were previously resolved with high-resolution synchrotron single-crystal X-ray diffraction.[32,33] The experimentally characterized aperiodic structures of $ScZn_{7.33}$ and $YbCd_{5.7}$ contain 37,483 and 56,155 atoms, within cubes of $(8\ nm)^3$ and $(10\ nm)^3$, respectively.

Figure 1a illustrates how we construct quasicrystal nanospheres for our DFT calculations. First, we select five random initial sites, each centered on a primary tetrahedral Tsai cluster. We then scoop out different-sized nanospheres around this center site, with nanoparticles ranging from 24 atoms (0.4 nm radius) to 740 atoms (1.41 nm). There are overlapping and partially occupied atomic sites in the XRD-refined structure files. To prepare ordered structures for DFT calculations, we deploy clustering and site-ordering algorithms (details discussed in Supplementary Information S1), which preserve the interatomic distances in Tsai-type icosahedral building-block motifs, while enforcing the overall quasicrystal stoichiometry.

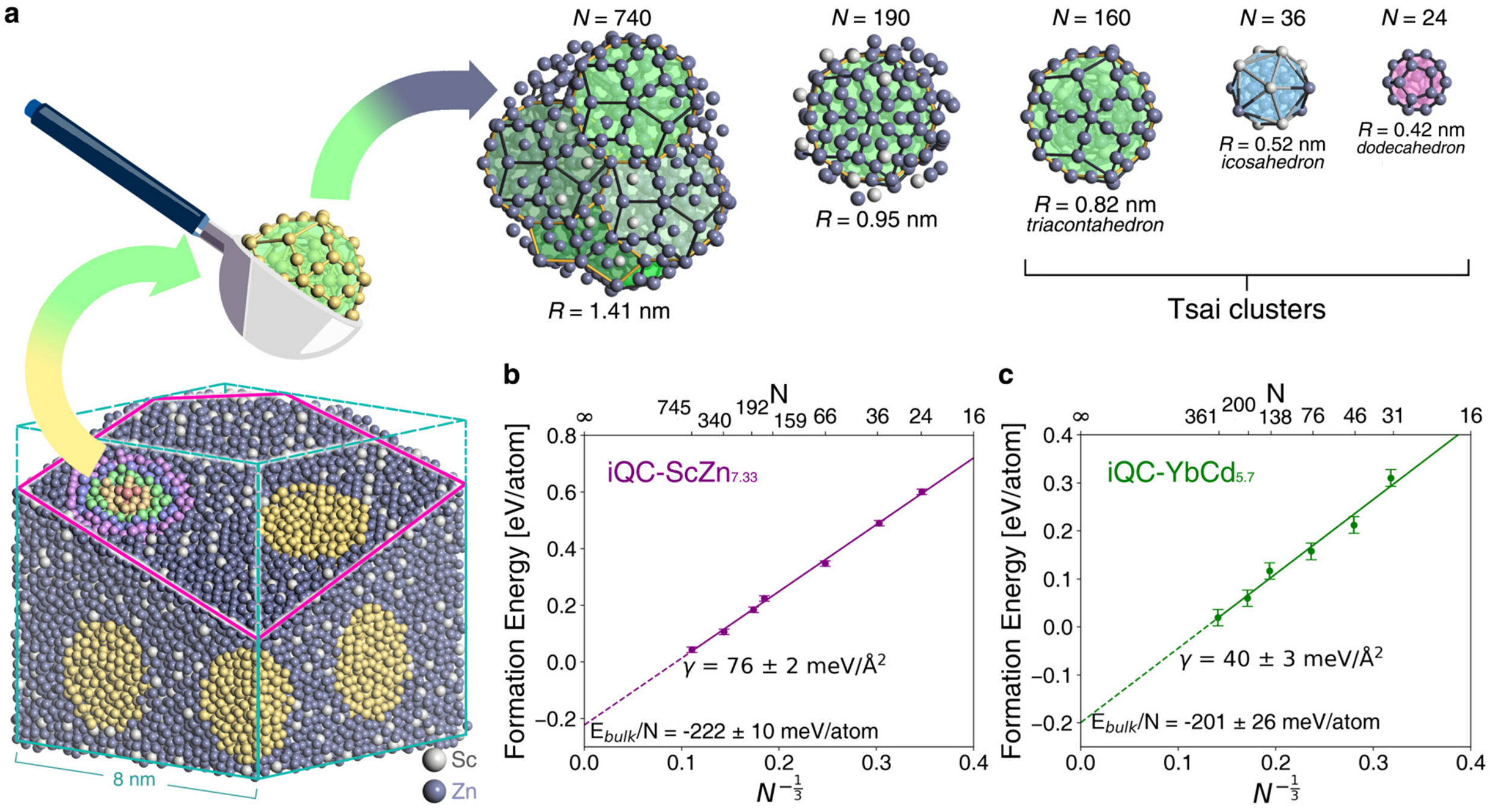

**Figure 1. Bulk and surface energies of icosahedral quasicrystals from scooped nanospheres. a,** Schematic illustration of scooping iQC-$ScZn_{7.33}$ nanospheres from the refined experimental atomistic structure. Yellow atoms represent random scooping sites. The right figure shows the series of nanoclusters with different sizes from scooping. **b,** The energies of iQC-$ScZn_{7.33}$ nanospheres show a linear relationship to $N^{-1/3}$. A linear regression can therefore be used to determine the bulk formation energy ($E_{bulk}$ / N) from the intercept, and surface energy ($\gamma$) from the slope. **c,** Calculated bulk formation and surface energy of iQC-$YbCd_{5.7}$.

To accurately extrapolate these nanoparticle energies to the bulk infinite limit, the biggest nanospheres should be as large as possible. However, the computational cost of plane-wave DFT calculations scale as $O(N^3)$, meaning nanoparticles with hundreds of atoms have considerable computational cost. Here we use the recently developed package DFT-FE,[34–36] which solves the Kohn-Sham equations using the discretized high-order finite-element method[37] in conjunction with Chebyshev filtered subspace iteration procedure[38] and leverage GPU acceleration.[39] This enables $O(N^2)$ scaling up to 30,000 electron systems (~1000 atoms of iQC-$ScZn_{7.33}$).[35] Convergence tests and benchmarking of DFT-FE formation energies are detailed in Supplementary Information S2.

DFT-FE further exhibits good parallel scaling, which enables structural relaxations on quasicrystal nanoparticles. We relax all atomic coordinates of our scooped nanospheres and provide the final relaxed structural coordinate files in the Supplementary Files. The ionic relaxations of the largest quasicrystal nanoparticles here required supercomputing resources at the frontier of exascale computing.[40,41] However, static calculations of our DFT-relaxed structures are reproducible on typical supercomputing resources.

Figure 1b and 1c show the DFT-calculated nanoparticle energies for iQC-$ScZn_{7.33}$ and iQC-$YbCd_{5.7}$ nanoclusters. We then fit a linear trend to $N^{-1/3}$, using the shape factor of a sphere $\eta = (36\pi)^{1/3}$. For iQC-$ScZn_{7.33}$, we calculate the bulk formation energy to be $E_{bulk} / N = -222 \pm 10$ meV/atom and the surface energy to be $\gamma = 76 \pm 2$ meV/Å$^2$. For iQC-$YbCd_{5.7}$, we calculate $E_{bulk} / N = -201 \pm 26$ meV/atom and $\gamma = 40 \pm 3$ meV/Å$^2$. The error bars arise from the standard deviation of the nanoparticle energies originating from the 5 different scooping sites, which samples different local aperiodic tiling environments and surface terminations for the different nanoparticles. In Supplementary Information S3, we show that our nano-scooping technique reproduces the energetics of crystalline Sc and $ScZn_6$, yielding bulk and surface energies that match traditional DFT calculations with periodic boundary conditions within 9 meV/atom and 3 meV/Å$^2$ for Sc, and 5 meV/atom and 2 meV/Å$^2$ for $ScZn_6$, respectively. This shows that the nano-scooping method is a reliable and accurate technique to evaluate quasicrystal bulk and surface energies.

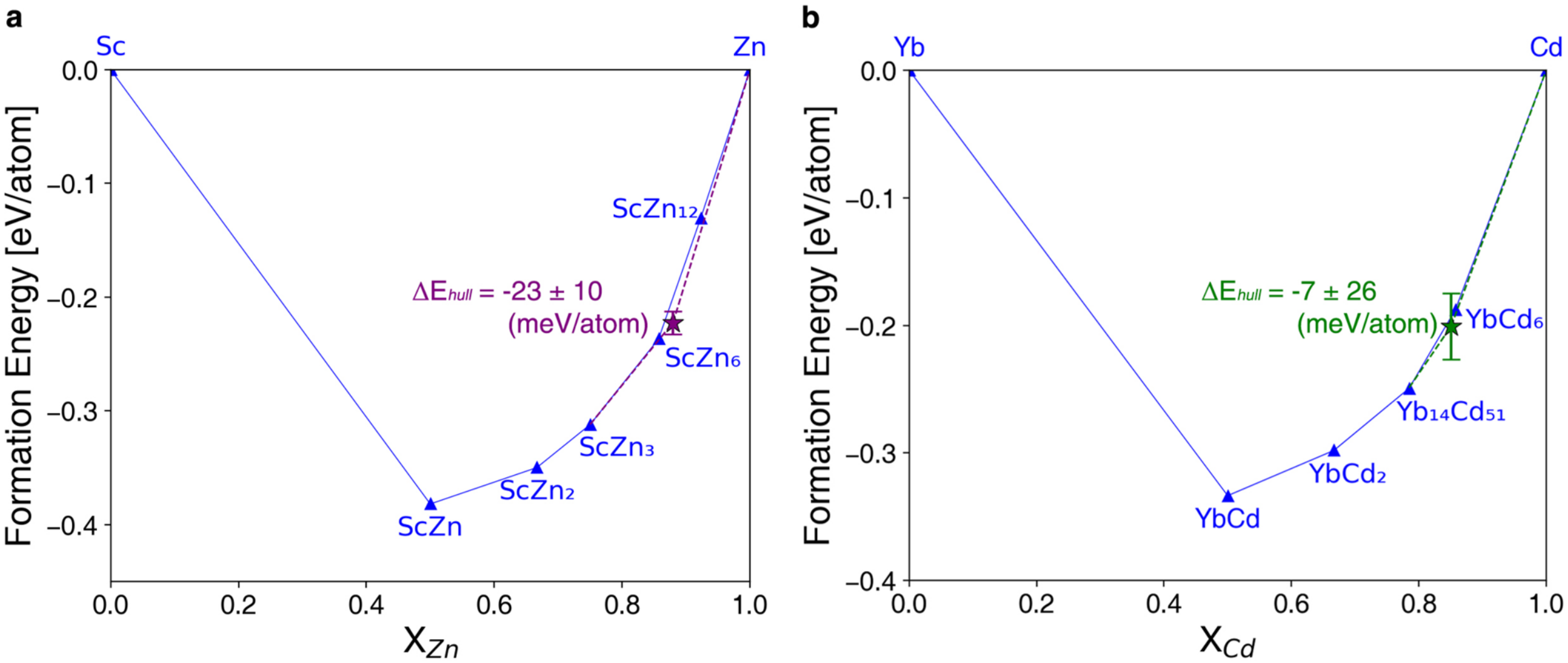

**Figure 2. DFT-calculated convex hull stability of icosahedral quasicrystals** for **a**, iQC-$ScZn_{7.33}$ and **b**, iQC-$YbCd_{5.7}$. The blue hulls are the plot of the most stable crystalline binary alloys connecting their calculated bulk formation energy from DFT.

Using these bulk formation energies, we next evaluate the $T = 0$ K convex hull stability of the icosahedral quasicrystal phases (Figure 2). For iQC-$ScZn_{7.33}$, the calculated bulk formation energy is below the existing Sc-Zn convex hull by $\Delta E_{hull} = -23 \pm 10$ meV/atom. In intermetallic systems, DFT predictions of ground-state structures are 96% accurate.[42] Therefore, based on our calculations, we assert with high confidence that the iQC-$ScZn_{7.33}$ quasicrystal is a thermodynamic ground-state phase. iQC-$ScZn_{7.33}$ was conjectured to be thermodynamically stable in the original experimental work, based on its thermal persistence under annealing at 390 °C for 22 h.[12] Our electronic structure calculations here provide first-principles validation of this original claim.

iQC-$YbCd_{5.7}$ was the first discovered binary icosahedral quasicrystal,[11] and was also argued to be stable due to it being a congruently melting phase—although this is an indirect argument as it could possibly be a eutectoid phase with low-temperature phase separation. We calculate the formation energy iQC-$YbCd_{5.7}$ to be 7 meV/atom below the existing convex hull. Although the error bar is ± 26 meV/atom, we solve that the entropy of iQC-$YbCd_{5.7}$ is lower than that of its competing $Yb_{14}Cd_{51}$ + $YbCd_6$ neighbors (Supplementary Information S4). If iQC-$YbCd_{5.7}$ were a metastable phase at low temperatures, we calculate that it would phase separate for temperatures below 561 K. Because there is no experimentally-reported phase separation upon cooling iQC-$YbCd_{5.7}$, this suggests that $YbCd_{5.7}$ is also most probably a $T = 0$ K ground state phase.

Our finding that the $ScZn_{7.33}$ and $YbCd_{5.7}$ quasicrystals are ground-state phases leads to a surprising conclusion that long-range translational symmetry is not necessary for the $T = 0$ K stability of inorganic solids. This result provokes deep and fundamental questions on the relationships between atomic structure and quantum chemical bonding in inorganic solids. Which short-range chemical bonds stabilize the triacontahedron building blocks, and why do long-range interactions between these triacontahedrons favor long-range aperiodicity? On the other hand—if aperiodic solids can be ground state phases, then what is so special about translational symmetry that enforces nearly all stable materials to be crystalline? Quasicrystals offer a gateway system through which we can interrogate these fundamental questions of solid-state chemistry.

## Competitive nucleation between intermetallics

When the Sc-Zn phase diagram was first thermodynamically assessed in 1997, five intermetallic phases were identified.[43] However, the icosahedral $ScZn_{7.33}$ quasicrystal was not discovered until 2010.[12] If iQC-$ScZn_{7.33}$ is thermodynamically stable, why did it take so long to discover? One mechanism may be that the other Sc-Zn intermetallics have lower surface energies than the quasicrystal, which could stabilize these competing intermetallics at the nanoscale.[30,31] All materials nucleate and grow through the nanoscale, meaning these competing intermetallics would dominate the nucleation kinetics[44] of solidification from the undercooled liquid—giving them a head start on subsequent crystal growth processes. Because we also calculated the surface energies of the iQC-$ScZn_{7.33}$ quasicrystal, we next evaluate if the other intermetallics in the Sc-Zn system can be nano-stabilized relative to the quasicrystal, and thereby be more kinetically favored to nucleate.

First, we use DFT to calculate surface energies for the other crystalline phases in the Sc-Zn system, considering high-index surface slabs with $\{hkl\}$ up to 3 (Supplementary Information S5).[45] Using these surface energies, in Figure 3a we extend the Sc-Zn convex hull along a third axis of Area-to-Volume ratio, or effectively, 1/R.[46] The slope of each intermetallic phase along the 1/R axis is $\gamma\eta/\rho$, corresponding to its surface energy, shape factor from the Wulff construction, and atomic density, respectively. Figure 3b shows an $E/N$ vs. 1/R slice of the size-dependent convex hull at the $ScZn_{7.33}$ composition. At this composition, iQC-$ScZn_{7.33}$ is the bulk equilibrium phase; however, the stoichiometric combination of $ScZn_6$ + $ScZn_{12}$ has a lower $\gamma\eta/\rho$ value than iQC-$ScZn_{7.33}$, such that $ScZn_6$ + $ScZn_{12}$ become nano-stabilized for $R < 0.63$ nm. The size-dependent phase diagram in Figure 3c indicates a strong likelihood of structural competition between quasicrystalline iQC-$ScZn_{7.33}$ and crystalline $ScZn_6$ and $ScZn_{12}$ during nucleation.

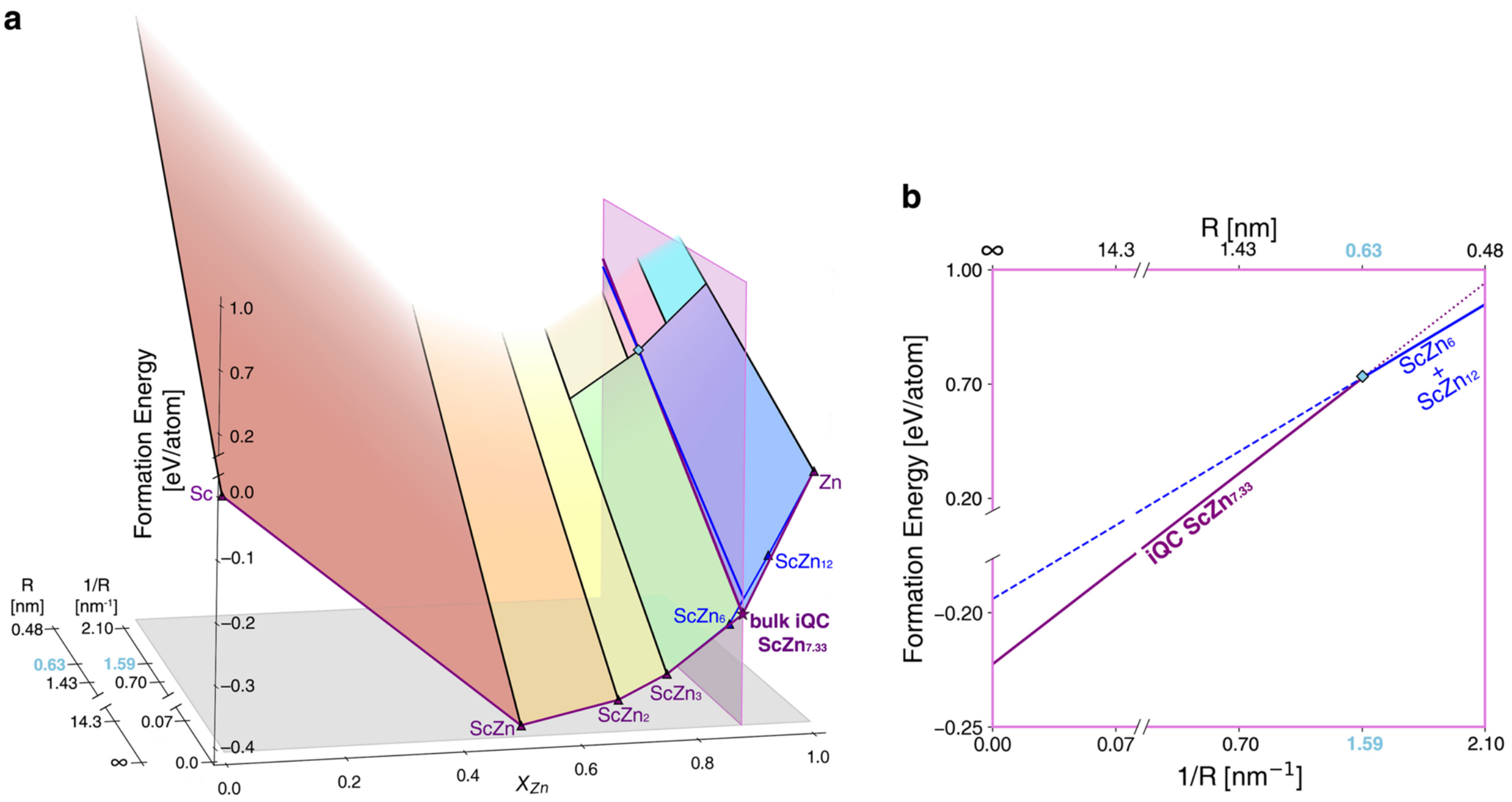

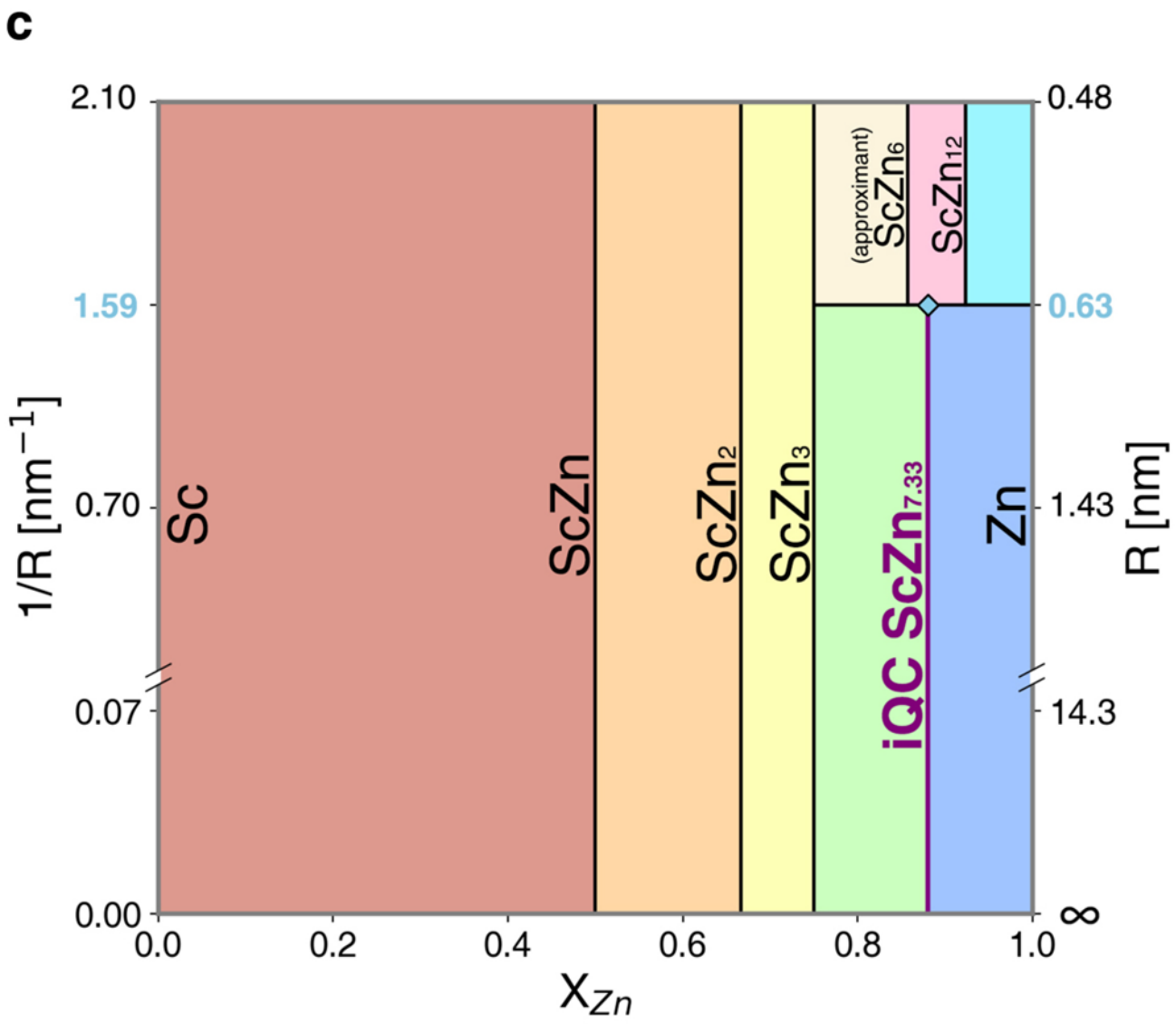

**Figure 3. Size-dependent stability of iQC-$ScZn_{7.33}$. a,** 3D convex hull of Sc-Zn system that includes a 1/R axis to represent surface area to volume ratio. The violet plane marks a composition slice at the $ScZn_{7.33}$ composition. **b,** 2D Slice energy diagram of the 3D convex hull in the Energy versus 1/R axis at the $ScZn_{7.33}$ composition (violet). The bold straight purple and blue lines represents the phase(s) with the lowest energy for a given 1/R. **c,** Projection of the 3D convex hull onto the size and composition axes (gray), producing a size-dependent phase diagram. Lines indicate pure compounds, and colored regions represent 2-phase coexistence.

During solidification, any intermetallic phase can nucleate from an undercooled melt, even if it has a different composition than the parent liquid phase. We next calculate the relative nucleation barriers for each solid, $\Delta G_c = 4/27\ [(\gamma\eta)^3 / (\Delta G_{\text{solid-liquid}})^2]$, where the driving force for solidification, $\Delta G_{\text{solid-liquid}}$, depends on the composition of the parent liquid, as well as the undercooling temperature.[47] To reference a Sc-Zn liquid free-energy to our DFT-calculated solid energies, we reassess the Sc-Zn phase diagram using the Extensible Self-optimizing Phase Equilibria Infrastructure (ESPEI) package.[48] We use our DFT-calculated $T = 0$ K convex hull for the bulk formation enthalpies of the solids, and we co-optimize the liquid free energy and solid entropies simultaneously, constrained by the experimentally-measured liquidus curve[43] and the congruent melting or peritectic decomposition condition of each intermetallic (details in Supplementary Information S6).[49] We then obtain $\mu_{\text{Sc,liquid}}$ and $\mu_{\text{Zn,liquid}}$ from the intercept rule on the tangent line of $G_{\text{liquid}}$ at each liquid composition and undercooling temperature—from which we evaluate $\Delta G_{\text{liquid–solid}} = G_{Solid} - \mu_{\text{Sc,liquid}} - \mu_{\text{Zn,liquid}}$ to each competing solid phase (Figure 4a).

Figure 4b plots the homogeneous nucleation barrier of each solid versus the undercooling temperature, starting from a liquid with the same composition of the quasicrystal ($x_{\text{Zn}}$= 0.88). At this liquid composition, the quasicrystalline approximant phase $ScZn_6$ has the lowest nucleation barrier for all undercooling temperatures. This is consistent with the experimental observations of Canfield *et al.*,[12] where $ScZn_6$ was found to be the primary phase in most solidification experiments. Only serendipitously did the quasicrystal present itself as small facetted icosahedral residues on the surfaces of $ScZn_6$ crystals, or on the walls of the crucible.[12]

To isolate single-grains of iQC-$ScZn_{7.33}$, Canfield *et al.* reported that they had to start from a very Zn-rich liquid with compositions ranging from $Sc_4Zn_{96}$ to $Sc_2Zn_{98}$.[12] In Figure 4c, we show the nucleation barrier of each phase when undercooled from a liquid of $x_{\text{Zn}}$ = 0.97 composition, which reveals that iQC-$ScZn_{7.33}$ becomes the lowest nucleation barrier phase below T < 603 K. This is because a higher driving force is needed to overcome the higher surface energy of iQC-$ScZn_{7.33}$ relative to $ScZn_6$, meaning a higher $\mu_{\text{Zn,liquid}}$ is needed. In Figure 4d, we repeat this nucleation analysis at all Sc-Zn liquid compositions—overlaying a 'kinetic phase diagram' on top of the thermodynamic Sc-Zn phase diagram, where the color shading corresponds to the phase with the lowest nucleation barrier for a given liquid composition and undercooling.

This mixed thermodynamic and kinetic phase diagram (Figure 4d) reveals key insights into the solidification dynamics of the Sc-Zn system. First, $ScZn_6$ is nucleation-preferenced across a broad region of the kinetic phase diagram, due to its comparatively low surface energy and high driving force. Once $ScZn_6$ nucleates, it can dominate the early stages of crystal growth. iQC-$ScZn_{7.33}$ is the ground-state phase, meaning it still has a propensity to nucleate at low temperatures, forming the small grains on the $ScZn_6$ crystals as reported experimentally. However, Figure 4d shows that iQC-$ScZn_{7.33}$ becomes nucleation-preferenced at high $x_{\text{Zn}}$ liquid compositions and small undercoolings. After it nucleates, iQC-$ScZn_{7.33}$ can be grown in the two-phase Liquid + iQC-$ScZn_{7.33}$ coexistence region—consistent with the experimentally-reported procedure to isolate phase-pure quasicrystals.[12] Figure 4 highlights the subtle interplay between thermodynamics and kinetics during materials synthesis, and illustrates why not all thermodynamically-stable phases are necessarily easy to synthesize.

We reproduce these size-dependent convex hulls and nucleation analyses for the Yb-Cd system in Supplementary Information S7. Because iQC-$YbCd_{5.7}$ is a congruently melting phase, it is the only phase that nucleates from the liquid below its congruent melting point.[50] This means it is fairly straightforward to obtain phase-pure iQC-$YbCd_{5.7}$, which plausibly explains why it was the first discovered binary icosahedral quasicrystal.[11]

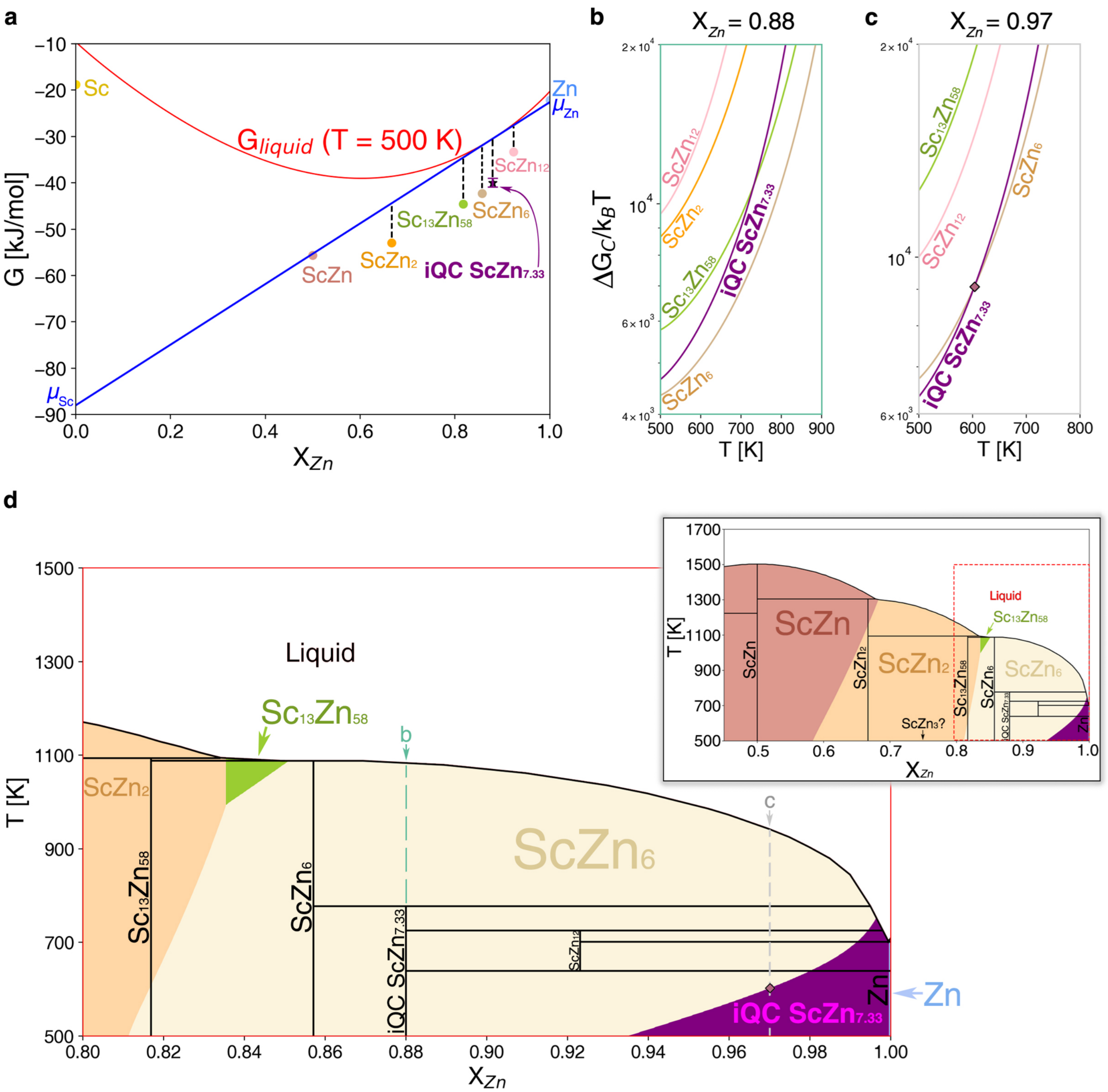

**Figure 4. Mixed thermodynamic and kinetic phase diagram of competitive nucleation in the Sc-Zn system. a,** Evaluation of the $\Delta G_{\text{liquid–solid}}$ from an undercooled liquid by the common tangent construction. The tangent line (blue) of $G_{\text{liquid}}$ (red) is drawn at iQC-$ScZn_{7.33}$ composition ($X_{Zn}$ = 0.88) and T = 500 K. The dashed black lines above the points correspond to $\Delta G_{\text{liquid–solid}}$ for each solid. **b,** Calculated free energy barrier of the critical nucleus ($\Delta G_C$) at quasicrystal composition ($X_{Zn}$ = 0.88) and **c,** $X_{Zn}$ = 0.97 with different temperature. **d,** Kinetic phase diagram of nucleation in Sc-Zn system. The black lines and compositions represent the equilibrium phase diagram calculated from ESPEI by referring DFT and experimental phase diagrams[12,43] and colored regions below the liquid indicate the phase having the lowest nucleation barrier at given temperature and composition. The vertical aquamarine and gray dashed line indicate the composition at $X_{Zn}$ = 0.88 and 0.97, respectively. The left figure zooms in on the regions marked with dashed red lines of the right figure.

## Outlook

Here, we presented a nanoparticle size-scaling technique to directly evaluate the bulk and surface energies of quasicrystals, overcoming a fundamental limitation of *ab initio* methods to compute the thermochemical properties of aperiodic matter. By opening the door to DFT calculations on quasicrystals, we can now unleash the full suite of first-principles methods to explore the fascinating *structure-chemistry-bonding* relationships of quasicrystalline materials.

For example, we can use finite-difference[51] or *ab initio* molecular dynamics methods[52] to elucidate the non-Debye nature of aperiodic phonons in quasicrystals. We can fit cluster expansions to construct partition functions for phason disorder,[53] which could reveal why some quasicrystals exhibit heat capacities up to 4.7 $k_B$/atom[54]—greatly exceeding the law of Dulong-Petit. Crystal Orbital Hamiltonian Population (COHP)[55] and Density Functional Theory-Chemical Pressure (DFT-CP) analysis[21,56,57] offers a pathway to understand which quantum-chemical bonds stabilize long-range aperiodic tilings. These DFT calculations can now be performed on real intermetallic quasicrystals using our technique, which will reveal how chemistry influences the various thermochemical properties of different quasicrystals. This computational method also encourages more experimental characterization of atomistic quasicrystalline structures, for which recently developed electron tomography methods may be promising.[58]

Our nanoparticle size-scaling approach can also be applied beyond quasicrystals—for example to partition the energetics of intermetallics with giant unit cells containing thousands of atoms (β-$Mg_2Al_3$, $NaCd_2$, *etc*),[59] or to other forms of poorly-crystalline matter including glasses, amorphous oxides, organic soft matter, and protein crystals.[60–62] Training machine-learned interatomic potentials[63] on first-principles calculations of these finite-sized nanoparticles may be a scalable pathway to DFT-derived insights into hierarchically aperiodic systems.

## Methods

### Structure refinement

Atomistic structures of Tsai-type $ScZn_{7.33}$ and $YbCd_{5.7}$ icosahedral quasicrystals (iQCs) were previously determined through high-resolution single-crystal X-ray diffraction experiments by Yamada et al.[33] and Hiroyuki et al.,[32] respectively. The site distribution of atoms was analyzed through the radial pair-correlation function of interatomic distance (Supplementary Information S1). Overlapping sites were eliminated by clustering the nearest atoms and averaging the distance and reanalyzed through partial and total radial pair-correlation function. Partially occupied sites of $ScZn_{7.33}$ were ordered by assigning atomic sites from a Gaussian distribution following experimentally measured stoichiometries.[32] The refined bulk structure files are in the Supplementary Files.

### Density functional theory

All electronic structure calculations for the intermetallics were performed using Density Functional Theory-Finite Element (DFT-FE) method[34,35] and validated against the Quantum Espresso (QE) package,[64,65] with the General Gradient Approximation (GGA), Perdew-Burke-Ernzerhof (PBE)[66] exchange functional, and Optimized Norm-Conserving Vanderbilt (ONCV)[67,68] pseudopotentials (Supplementary Information S2). The optimal *k*-points for all periodic crystalline alloys were generated from automatic density in the Pymatgen package.[69] The discretization for QE and DFT-FE are chosen such that the discretization errors in QE and DFT-FE are both below $10^{-6}$ Ha/atom range. The chemical accuracy, ionic force, and cell stress convergence threshold for geometry relaxation of crystalline alloys were set to $10^{-5}$ Ha/atom, $10^{-4}$ Ha/Bohr, and $10^{-6}$ Ha/Bohr$^3$, respectively. The Fermi-Dirac smearing temperature was set to 500 K. For the finite-size calculations of scooped nanospheres, gamma-point calculations were employed in DFT-FE. The ionic force threshold was set to less than $9 \times 10^{-4}$ Ha/Bohr for structure relaxation.

### Surface energy calculations of crystalline compounds

Surface energies of the crystalline intermetallic compounds were calculated from slab structures, generated by cleaving bulk phases using the Pymatgen package.[45] The maximum Miller index was set to 3 and minimal vacuum length normal to the surface was set to 15 Å. Optimal atomic length was selected from the surface energy convergence test by thickness for selected planes (Supplementary Information S5). The inclination of stable slab planes was searched from 'WulffShape' in the Pymatgen package[70] for $\gamma$-plot construction. The weighted surface energy was calculated from the weighted mean of the surface energy by the area of the Wulff shape planes.

### Phase diagram assessment

Gibbs free energies for the solid and liquid phases were thermodynamically assessed using the Extensible Self-optimized Phase Equilibria Infrastructure (ESPEI) package.[48] The pure elemental solid and liquid free energy parameters were taken from the SGTE database.[71] ESPEI generates model parameters incorporating the published experimental liquidus curve[12,43,50] and our DFT formation energies of crystal structures including the quasicrystals. The non-ideal mixing energies were calculated from Redlich-Kister model[72] based on the thermochemical data using Muggianu extrapolation.[73] Interaction parameters and formation entropies of intermetallic compounds were simultaneously optimized via the Bayesian ensemble Markov Chain Monte Carlo (MCMC) method[74] employing triangular priors ranging from $\pm 0.8\ \theta$ where $\theta$ represents the initial parameter. The detailed optimization procedure and assessment are described in Supplementary Information S6.

**Acknowledgements:** W. B. thanks Dr. H. Takakura and T. Yamada for providing atomistic structures of $ScZn_{7.33}$ and $YbCd_{5.7}$ quasicrystals.

**Funding:** The work by W. B., S. T., and W. S. was supported by the U.S. Department of Energy (DOE), Office of Science, Basic Energy Sciences (BES), under Award No. DE-SC0021130. V.G. and S.D. thank the support of U.S. Department of Energy Basic Energy Sciences (DE-SC0008637) that supported the development of DFT-FE. The author acknowledges the Texas Advanced Computing Center (TACC) at the University of Texas at Austin, using grant award TG-MAT210016 from Extreme Science and Engineering Discovery Environment (XSEDE) and Advanced Cyberinfrastructure Coordination Ecosystem: Services & Support (ACCESS), National Energy Research Scientific Computing (NERSC) center at Lawrence Berkeley National Laboratory, which is supported by the Office of Science of the U.S. Department of Energy under Contract No. DE-AC02-05CH11231 using award BES-ERCAP0020148, and Oak Ridge Leadership Computing Facility at the Oak Ridge National Laboratory, which is supported by the Office of Science of the U.S. Department of Energy under Contract No. DE-AC05-00OR22725 for providing high-performance computing resources that have contributed to the research results.

**Author contributions:** Conceptualization: W. B. and W. S. Methodology: W. B., S. D., S. T., V. G., and W. S. Investigation: W. B., S. D., and S. T. Visualization: W. B. and S. T. Funding acquisition: W. S. and V. G. Supervision: W. S. Writing – original draft: W. B., S. T., and W. S. Writing – review and editing: W. B., S. D., S. T., V. G., and W. S.

**Competing interests:** We declare no competing interests.

**Data and materials availability:** All data are available in the manuscript or the Supplementary Information.

# Quasicrystal bulk and surface energies from density functional theory

## Supplementary Information

Woohyeon Baek[1], Sambit Das[2], Shibo Tan[1], Vikram Gavini[1,2], Wenhao Sun[1*]

[1]Department of Materials Science and Engineering, University of Michigan, Ann Arbor, MI, USA

[2]Department of Mechanical Engineering, University of Michigan, Ann Arbor, MI, USA

**Contents**

## S1. Structure refinement of Tsai-type iQC-$ScZn_{7.33}$ and $YbCd_{5.7}$ raw structure data

The atomistic structure data of Tsai-type $ScZn_{7.33}$ and $YbCd_{5.7}$ icosahedral quasicrystals (iQCs) that were determined through high-resolution single-crystal X-ray diffraction experiments by Yamada et al.[1] and Hiroyuki et al.,[2] respectively were used in this study. The Tsai-type iQC consists of three building units: the Tsai cluster (TC), were used in this study. The Tsai-type iQC consists of three building units: the Tsai cluster (TC), the obtuse rhombohedron (OR), and the acute rhombohedron (AR). The TC is a series of five consecutive polyhedra arranged like onion shells: tetrahedron, dodecahedron, icosahedron, icosidodecahedron, and rhombic triacontahedron. Each TC is aggregated by OR or AR linkers, which make up the entire atomic structure. The linked local structure resembles two approximant (derivative) crystals. The 1/1 approximant[3,4] is a body-centered cubic structure in which TCs are linked by sharing a face along two-fold directions or OR along three-fold directions. The 2/1 approximant[5,6] forms a staggered stacking of TCs by introducing diagonal AR linkage.

The discrepancy between global symmetry operations and local symmetry breaking leads to slight deviations in atomic coordination from the ideal geometry.[7] The innermost tetrahedron of the TC breaks icosahedral symmetry, distorting neighboring clusters.[8] Additionally, chemical disorder affects overlapping and partially occupied sites.[1] The iQC-$ScZn_{7.33}$ experiences distortion in double Friauf polyhedron (DFP) sites due to partial substitutions of Sc by Zn. Tetrahedron sites can rapidly reorient their atomic configuration inside the clusters within a few picoseconds,[9] leaving a cluster of partially-occupied overlapping atomic sites, which must be refined to avoid repulsion errors in DFT calculations.

The atomic positions and local structure of iQC-$ScZn_{7.33}$ and $YbCd_{5.7}$ were analyzed using the indexed positions of iQC-$ScZn_{7.33}$ determined by Yamada et al.[1] (**Figure S1**), through the radial pair-correlation function. The pair-correlation density profile of the raw structure shows strong peaks at interatomic distances less than 1 Å, indicating that the positions of atoms were almost overlapping each other (**Figure S2**). Analysis of the partial radial distribution by sites revealed that most overlapping atoms are in tetrahedron sites (**Figure S3** ⓘ). All overlapping sites including tetrahedron sites were reduced by clustering the nearest atoms and averaging the distance (**Figure S4a**). The density profile after the reduction shows almost no peaks at distances less than 2 Å, while preserving the original geometry of the clusters (**Figure S2 and S3**).

The iQC-$ScZn_{7.33}$ has partially occupied of Sc and Zn on icosahedron (**Figure S1** ⓙ) and prolate AR (DFP) sites locate around TCs (**Figure S1** ⓖ). These partially occupied sites were refined by randomly assigning atomic sites from a Gaussian distribution, based on the experimentally measured Sc/Zn ratio (0.74/0.26 for icosahedron and 0.48/0.52 for DFP site) (**Figure S4b**).

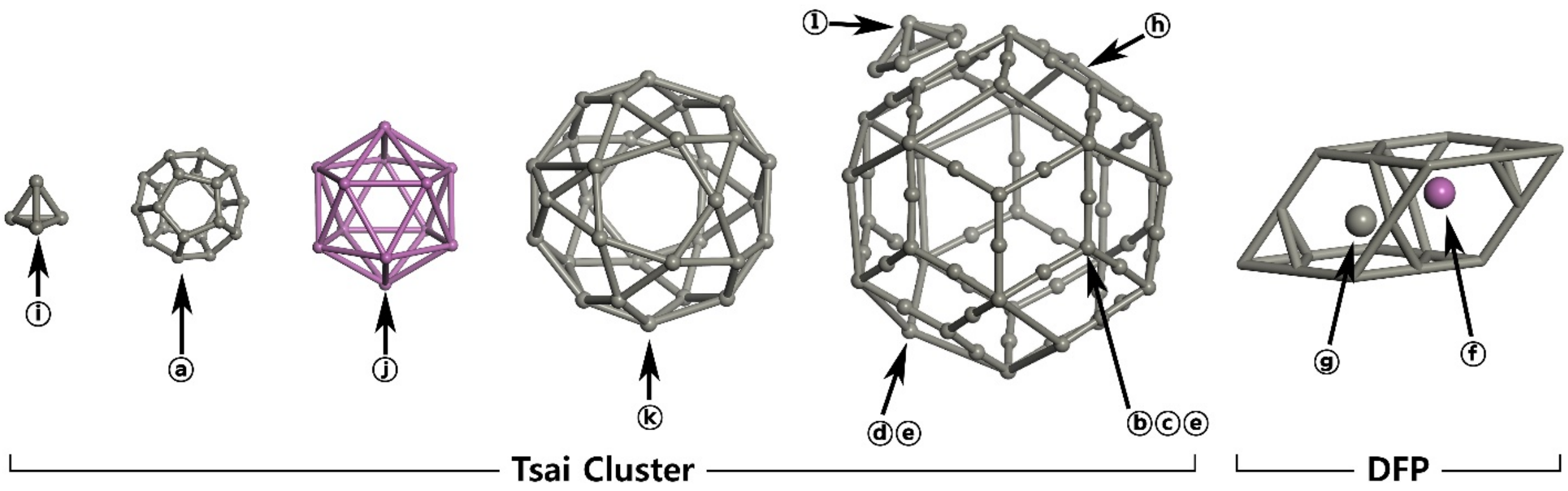

**Figure S1.** Atomic position indices of Tsai-type iQC TC and DFP.

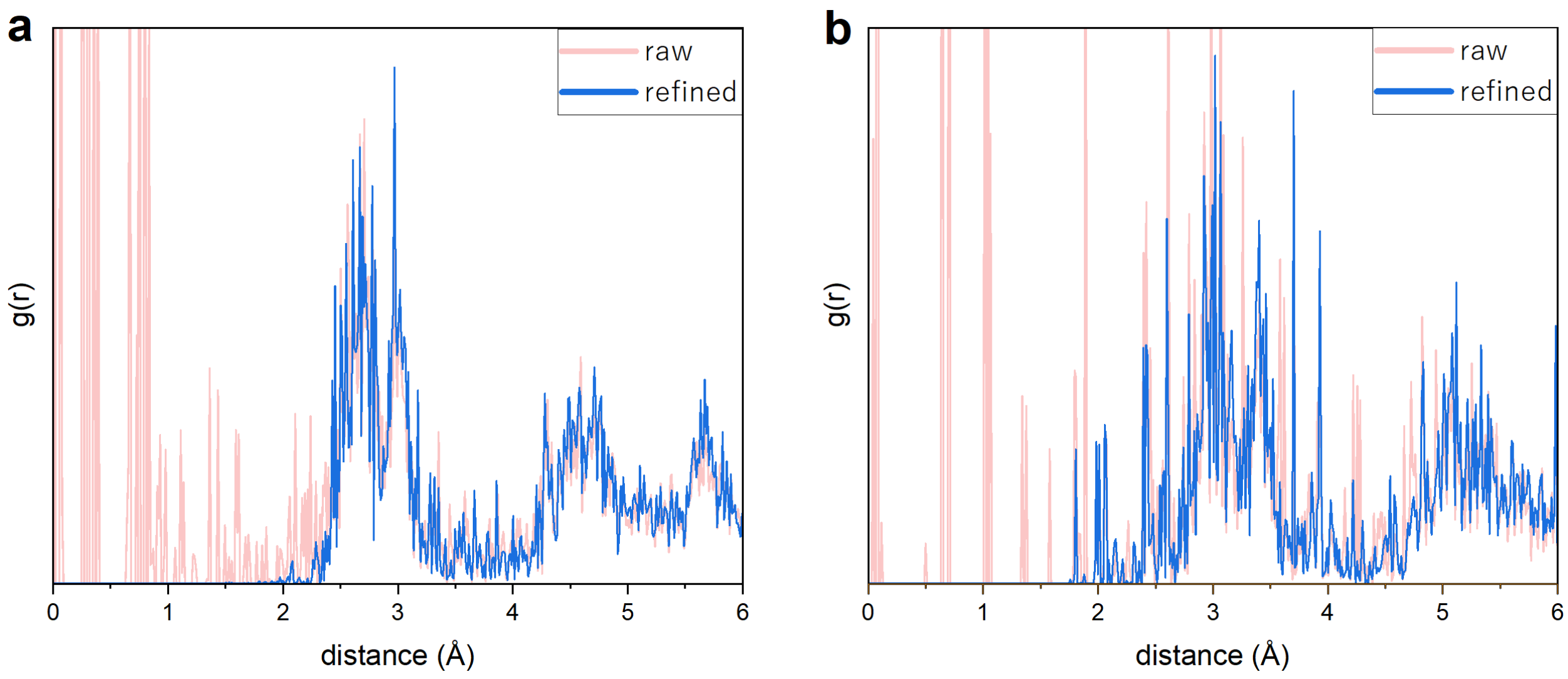

**Figure S2.** Radial pair-correlation density profiles for experimentally reported atomistic structures (red) and structurally refined corrections (blue) for **a,** iQC-$ScZn_{7.33}$ and **b,** iQC-$YbCd_{5.7}$ quasicrystals.

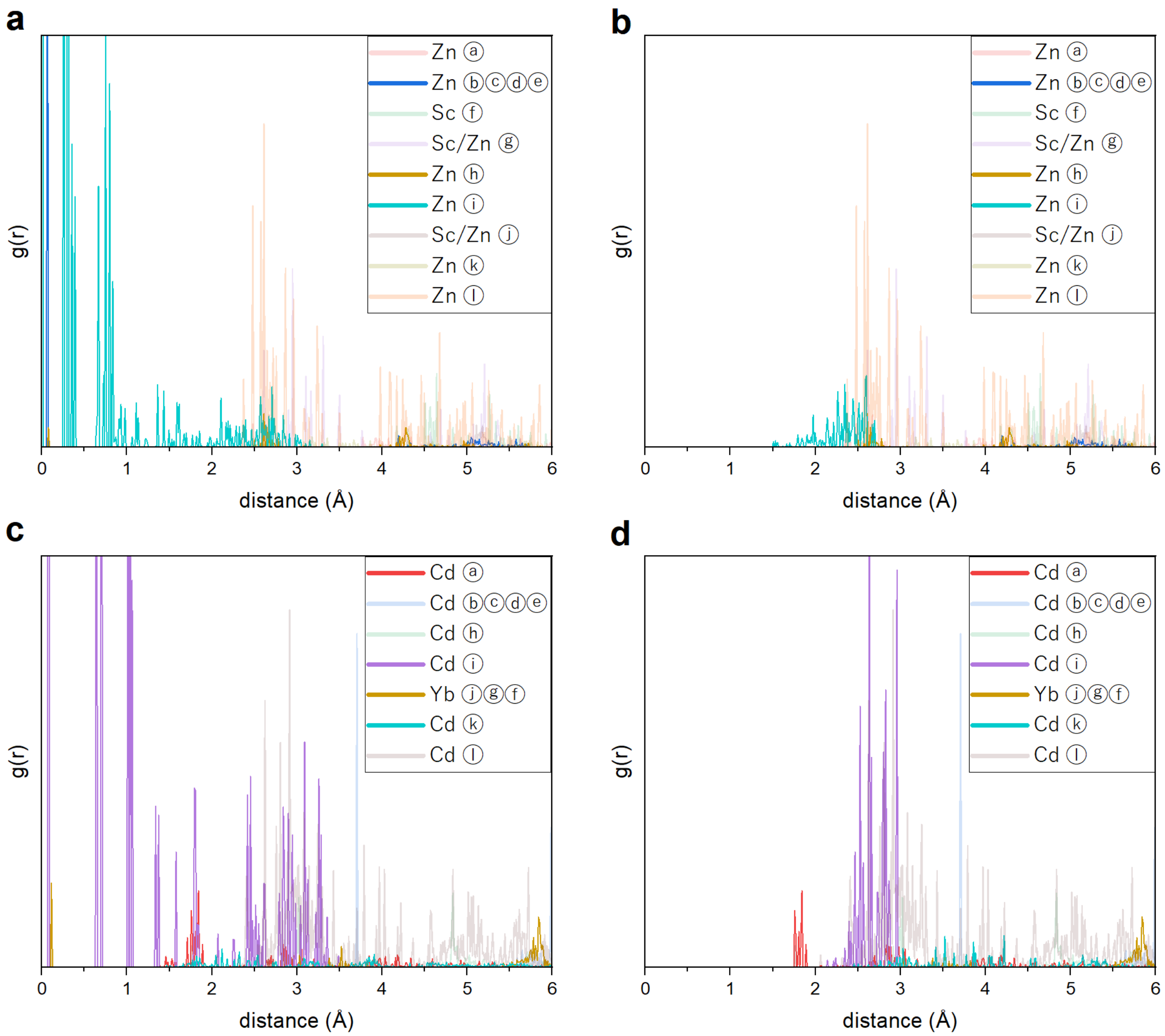

**Figure S3.** Partial radial pair-correlation density profile of **a,** raw and **b,** refined iQC-$ScZn_{7.33}$, and **c,** raw and **d,** refined iQC-$YbCd_{5.7}$ atomic structure data. The circled alphabet notation refers to atomic sites of Tsai-type iQC (**Figure S1**). Sites containing atoms closer than around 1.5 Å interatomic distance are highlighted opaquely.

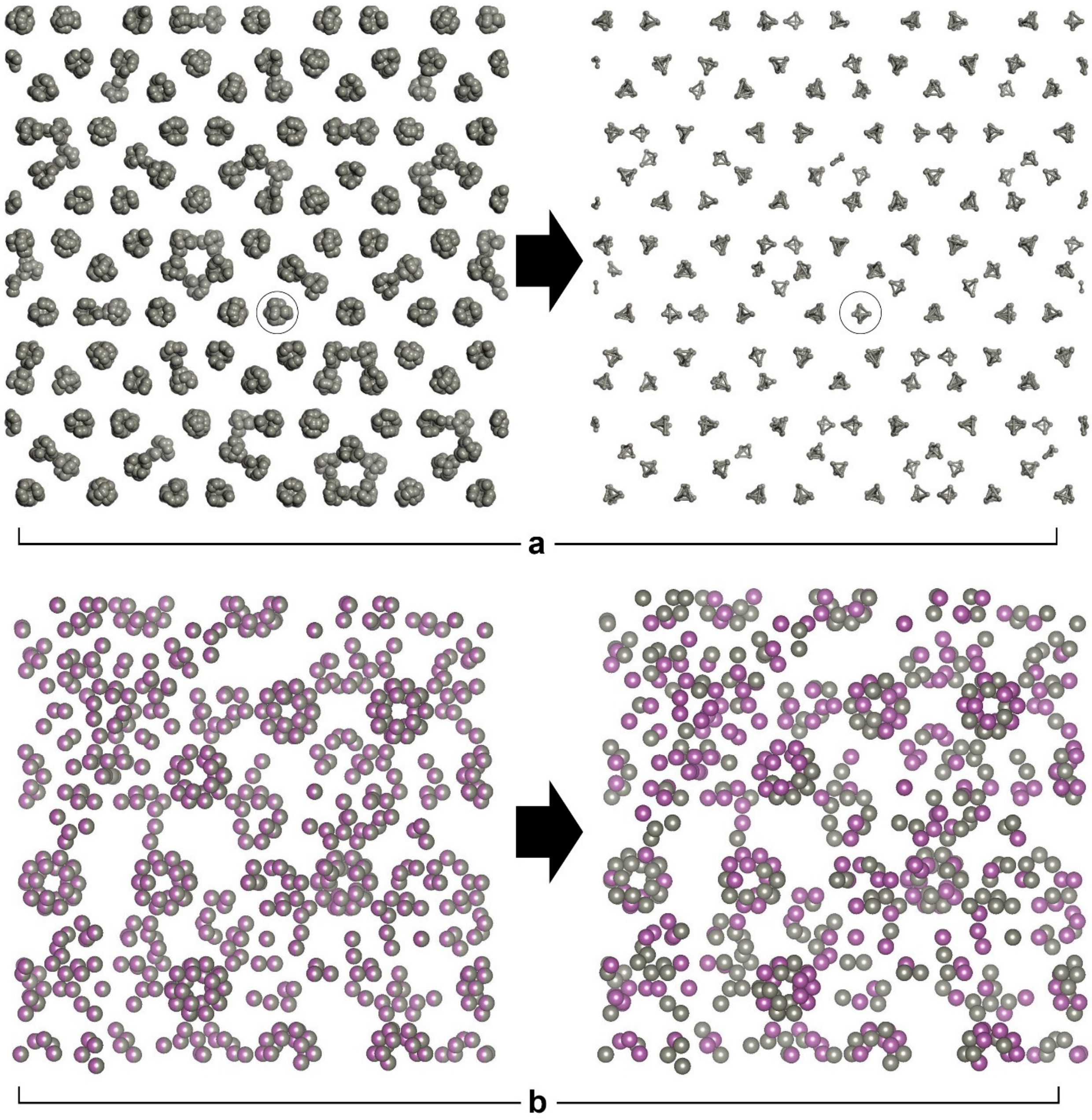

**Figure S4.** Schematic illustration of structure refinement. **a,** The reduction of overlapping atoms in tetrahedron sites (ⓘ) by clustering and **b,** partial occupation sites of DFP sites (ⓖ) from raw (left) to refined (right) atomic structures.

## S2. Density functional theory

All electronic structure calculations for the intermetallics were performed using Density Functional Theory-Finite Element (DFT-FE) method[10,11] and validated against the Quantum Espresso (QE) package,[12,13] with the General Gradient Approximation (GGA), Perdew-Burke-Ernzerhof (PBE)[14] exchange functional, and Optimized Norm-Conserving Vanderbilt (ONCV)[15,16] pseudopotentials. The optimal *k*-points for all periodic crystalline alloys were generated from automatic density in the Pymatgen package.[17] The discretization for QE and DFT-FE are chosen such that the discretization errors in QE and DFT-FE are both below $10^{-6}$ Ha/atom range. The chemical accuracy, ionic force, and cell stress convergence threshold for geometry relaxation of crystalline alloys were set to $10^{-5}$ Ha/atom, $10^{-4}$ Ha/Bohr, and $10^{-6}$ Ha/Bohr$^3$, respectively. The Fermi-Dirac smearing temperature was set to 500 K.

The DFT calculations for crystalline structures reproduces the same formation energies and convex hull when calculated from DFT-FE and compared with QE (**Figure S5**). This validates that the DFT-FE implementation has quantitative agreement with conventional crystalline DFT calculations. For the finite-size calculations of scooped nanospheres, gamma-point calculations were employed in DFT-FE. The ionic force threshold was set to less than $9 \times 10^{-4}$ Ha/Bohr for structure relaxation.

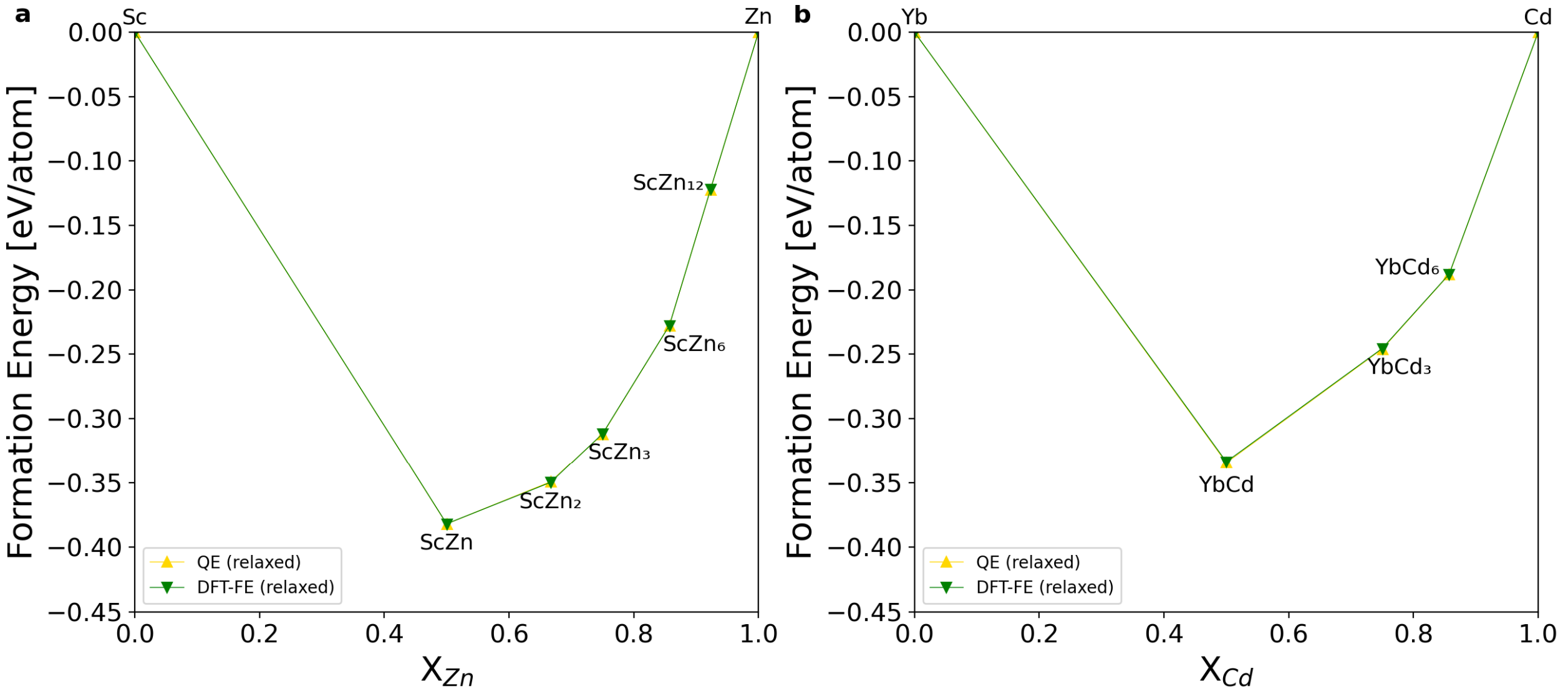

**Figure S5.** Formation energy of structurally static and relaxed QE and DFT-FE simulations for selected phases in **a,** $Sc_xZn_y$ and **b,** $Yb_xCd_y$ binary alloy systems.

The formation energy of a crystalline compound and scooped nanocluster can be calculated from DFT simulations.

$$\frac{E_{NP}}{N} = \frac{E_{A_xB_{(1-x)}} - \left[N_A E_A + N_B E_B\right]}{N}$$

where $E_{AxB(1-x)}$ is calculated total atomic energy of nanoparticles from DFT, $N_A$ and $N_B$ are the number of atoms of A (Sc, Yb) and B elements (Zn, Cd) that satisfies $N_A + N_B = N$, and $E_A$ and $E_B$ are the atomic energy of A and B elements, respectively.

**S3. Calculation of bulk formation and surface energies of crystalline phases from scooped nanoclusters**

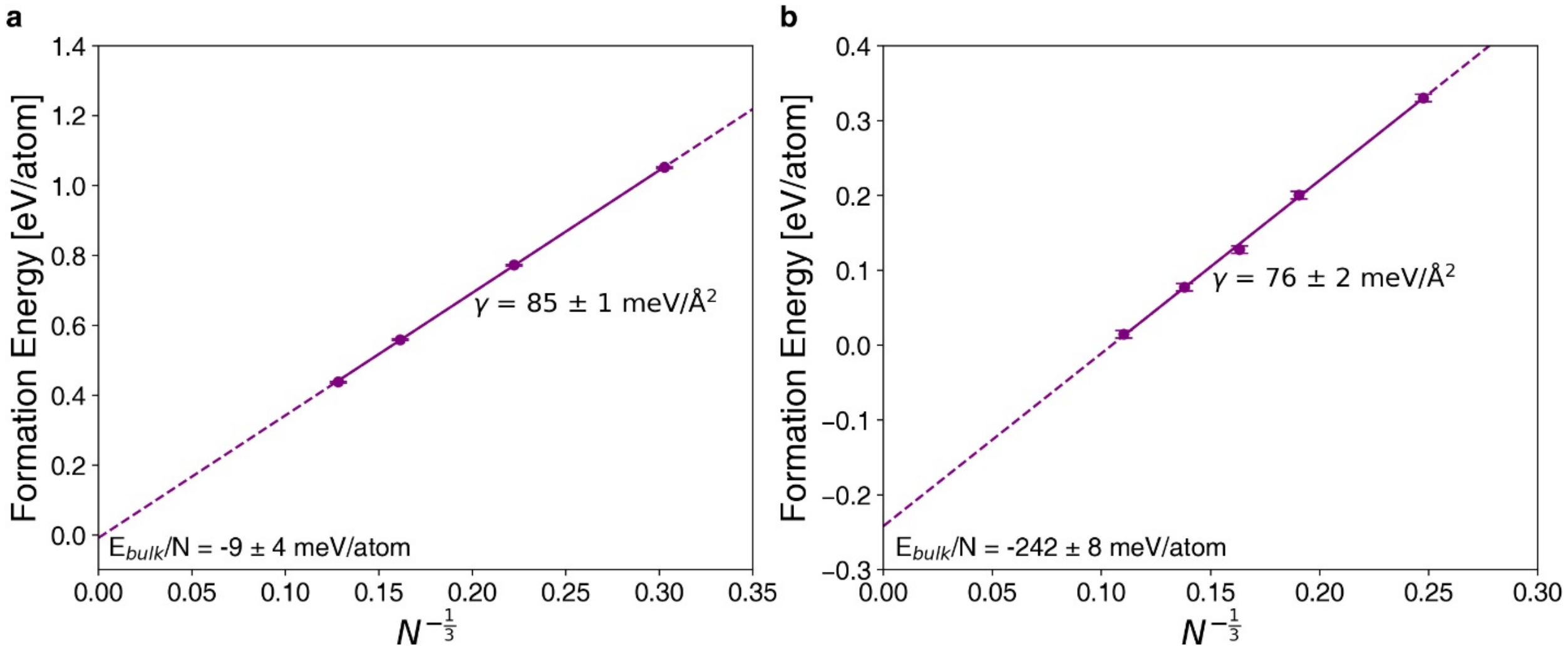

**Figure S6.** Calculated bulk formation and surface energy of **a,** Sc and **b,** $ScZn_6$ from linear regression of nanocluster formation energy to $N^{-1/3}$.

To confirm that the nano-scooping method we applied to quasicrystals is valid, we used the same technique on Sc and $ScZn_6$, to check if the calculated bulk and surface energies agree with traditional DFT calculations using periodic boundary conditions. We scooped a series of nanospheres with different numbers of atoms from 36 to 475 for Sc and from 66 to 747 for $ScZn_6$. The formation energies of scooped nanoclusters are calculated from relaxed structures. **Figure S6** shows a linear trend of calculated formation energy to $N^{-1/3}$ of nanospheres. From the linear regression, we calculate the bulk energies and surface energies for Sc to be $E_{bulk} / N = -9 \pm 4$ meV/atom and $\gamma = 85 \pm 1$ meV/Å$^2$, and for $ScZn_6$ we calculate $E_{bulk} / N = -242 \pm 8$ meV/atom and $\gamma = 76 \pm 2$ meV/Å$^2$. The periodic DFT simulations of Sc and $ScZn_6$ calculates the bulk formation and surface energy to $E_{bulk} / N = 0$ meV/atom (by definition, since Sc is an elemental reference state) and $\gamma = 82$ meV/Å$^2$ for Sc, and $E_{bulk} / N = -237$ meV/atom and $\gamma = 78$ meV/Å$^2$ for $ScZn_6$ (**Table S3**). The energy difference of the bulk formation energy is 9 and 5 meV/atom and surface energy is 3 and 2 meV/Å$^2$ for Sc and $ScZn_6$, respectively.

**S4. Gibbs free energy of Yb-Cd system at iQC-$YbCd_{5.7}$ composition**

Here, we assess the $T = 0$K stability iQC-$YbCd_{5.7}$ within the context of its DFT-FE calculated error bars. To do so, we assess the liquid free energy of the Yb-Cd system using the CALPHAD (CALculation of PHAse Diagram) approach. Given the melting points of iQC-$YbCd_{5.7}$, $Yb_{14}Cd_{51}$, and $YbCd_6$, we can fit their linearized entropies. This enables us to evaluate the scenario where if iQC-$YbCd_{5.7}$ is indeed metastable and has a low-temperature phase separation, what its critical eutectoid temperature should be.

The Gibbs free energy of liquid and solid compounds is calculated from the following equation.

$$G_{Yb_xCd_y} = \Delta_f H^{\varphi} - TS_{Yb_xCd_y}$$

where $\Delta_f H^{\varphi}$ is enthalpy of formation of the stochiometric phase φ (**Supplementary Information S6**), T is temperature, and $\Delta S_{Yb_xCd_y}$ is calculated entropy of $Yb_xCd_y$. The absolute entropy of compounds is calculated based on a thermodynamic definition that the enthalpy of elements is 0 at ambient condition.

$$S_{Yb_xCd_y} = \frac{\Delta_f H^{\varphi} - G_L^{Yb_xCd_y}(T_m^{Yb_xCd_y})}{T_m^{Yb_xCd_y} - 298.15}$$

where $G_L^{Yb_xCd_y}$ is calculated Gibbs free energy of liquid from CALPHAD optimization (**Supplementary Information S6**) and $T_m^{Yb_xCd_y}$ is melting temperature of $Yb_xCd_y$.

The calculated entropy for iQC-$YbCd_{5.7}$ and $Yb_{14}Cd_{51}$ + $YbCd_6$ is 98.527 and 98.912 J/mol·K, respectively. The upper-bound of the error bar for the formation energy of iQC-$YbCd_{5.7}$ increases its enthalpy to be above the hull. If iQC-$YbCd_{5.7}$ lies on the hull, the enthalpy would be -18,680 J/mol and the entropy would increase to 99.635 J/mol·K. The phase transition from iQC-$YbCd_{5.7}$ to $Yb_{14}Cd_{51}$ + $YbCd_6$ can then be calculated to occur at T < 561.5 K (Fig. S12A). If the enthalpy of iQC-$YbCd_{5.7}$ goes above the hull, the phase transition temperature also increases (Fig. S12B). However, the transition temperature then becomes close to or even exceeding the melting temperature of $YbCd_{5.7}$ (T = 909 K), as its enthalpy is above the hull. The previous experiment observed that the quasicrystalline $YbCd_{5.7}$ phase solidifies in a fully annealed single-crystal state.[18] Therefore, the Gibbs free energy of iQC-$YbCd_{5.7}$ should be lower than that of $Yb_{14}Cd_{51}$ + $YbCd_6$ at low temperature, and the formation enthalpy of iQC-$YbCd_{5.7}$ should be below the convex hull of competing crystalline intermetallic phases.

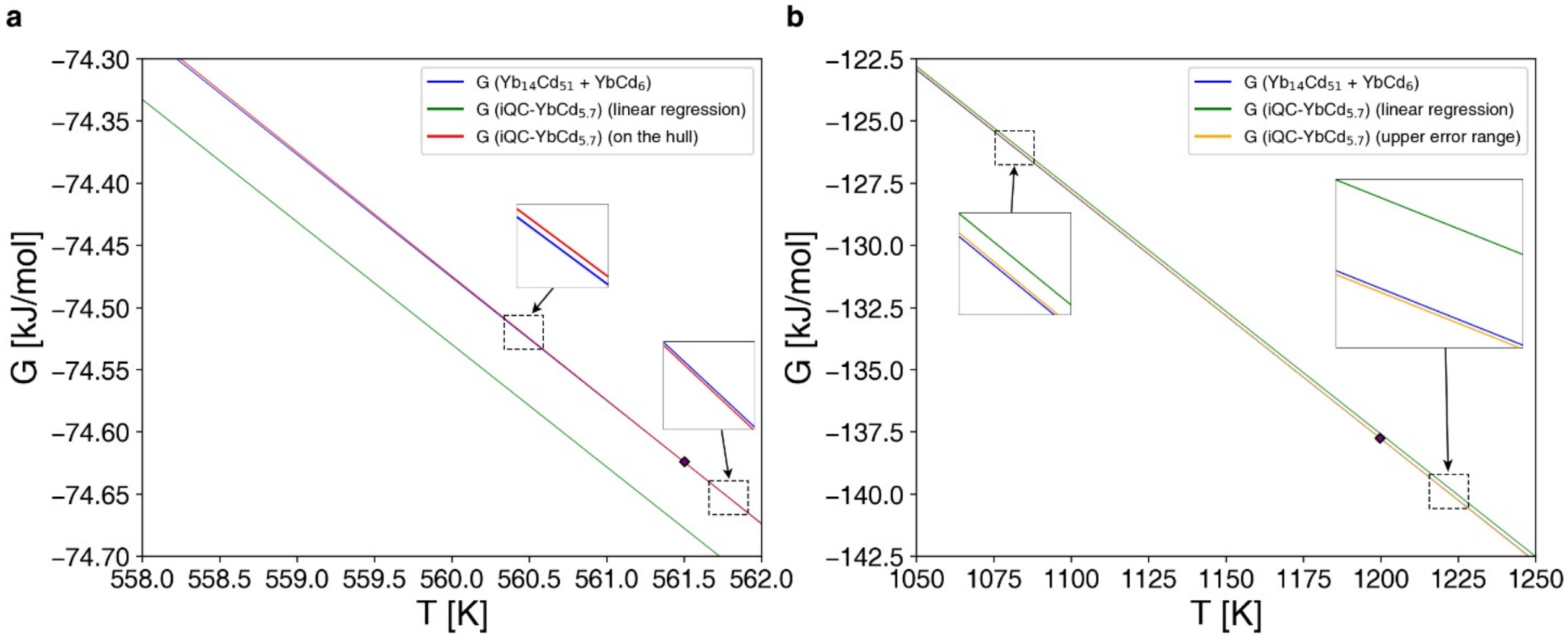

**Figure S7.** Calculated Gibbs free energy diagram at iQC-$YbCd_{5.7}$ composition ($X_{Cd}$ = 0.85) when the bulk formation energy is **a,** on the hull and **b,** the upper limit of the error above the hull. The blue line is $Yb_{14}Cd_{51}$ + $YbCd_6$, red is iQC-$YbCd_{5.7}$ if the formation energy is on the hull, orange is iQC-$YbCd_{5.7}$ if the formation energy is the upper limit of the error, and green is iQC-$YbCd_{5.7}$ from the intercept of linear regression (**Figure 1c**). The purple diamond is the intersection point of $Yb_{14}Cd_{51}$ + $YbCd_6$ and iQC-$YbCd_{5.7}$.

## S5. Wulff construction of crystalline alloys

The Wulff construction of all crystalline alloys is built by calculating the surface energy of cleaved faces.

$$\gamma = \frac{E_{slab} - N_{slab}\left(\frac{E_{bulk}}{N_{bulk}}\right)}{2A}$$

where $E_{slab}$ is the energy and $N_{slab}$ is the total number of atoms in a slab cell, and $E_{bulk}$ and $N_{bulk}$ are in bulk periodic cell. The slab structures were generated by cleaving bulk phases from 'generate all slabs' in the Pymatgen package.[19] The maximum Miller index was set to 3 and minimal vacuum length normal to the surface was set to 15 Å. Optimal atomic length was selected from the surface energy convergence test by thickness for selected planes (**Figure S8**). The inclination of stable slab planes was searched from 'WulffShape' in the Pymatgen package[20] for γ-plot construction (**Table S1 and S2**, **Figure S9 and S10**). **Table S3 and S4** summarize all calculated parameters of $Sc_xZn_y$ and $Yb_xCd_y$ alloys from the Wulff construction, respectively.

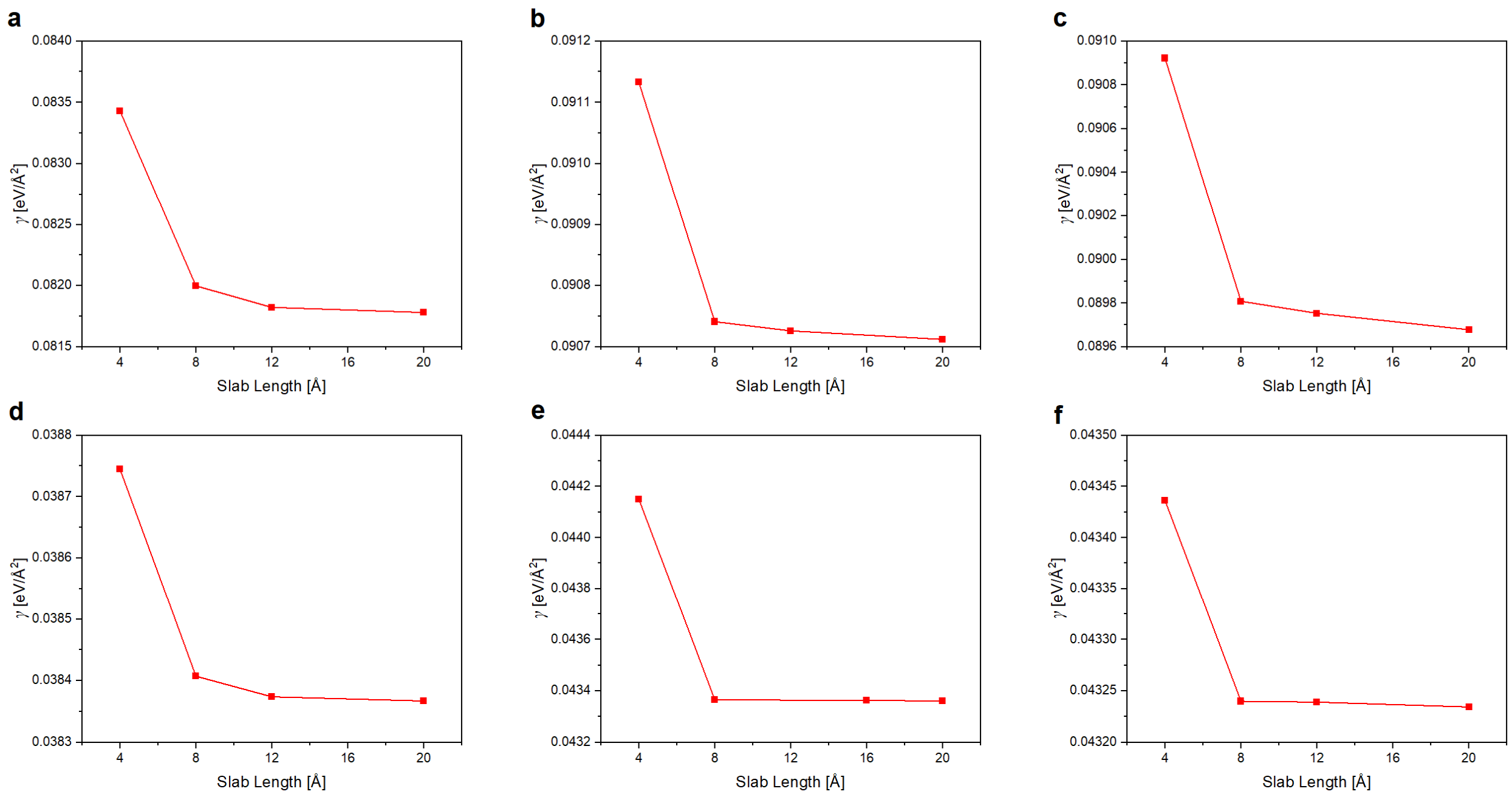

**Figure S8.** Convergence test of slab calculation by atomic thickness for **a,** (110), **b,** (211), and **c,** (321) ScZn planes, and **d,** (110), **e,** (211), and **f,** (321) YbCd planes.

**Sc**

| plane | γ [eV/Å$^2$] | γ [J/m$^2$] |
|---|---|---|
| (0, 0, 1) | 0.079 | 1.259 |
| (1, 0, 0) | 0.099 | 1.587 |
| (1, 0, 1) | 0.078 | 1.244 |
| (1, 0, 2) | 0.085 | 1.369 |
| (1, 0, 3) | 0.084 | 1.349 |
| (1, 1, 0) | 0.080 | 1.288 |
| (1, 1, 1) | 0.089 | 1.421 |
| (2, 0, 1) | 0.091 | 1.451 |
| (2, 0, 3) | 0.080 | 1.280 |
| (2, -1, 2) | 0.085 | 1.367 |
| (2, -1, 3) | 0.086 | 1.377 |
| (2, 2, 1) | 0.086 | 1.371 |
| (2, 2, 3) | 0.087 | 1.401 |
| (3, 0, 1) | 0.079 | 1.271 |
| (3, 0, 2) | 0.079 | 1.270 |
| (3, -1, 0) | 0.092 | 1.467 |
| (3, 1, 0) | 0.081 | 1.293 |
| (3, -1, 1) | 0.085 | 1.363 |
| (3, 1, 1) | 0.091 | 1.455 |
| (3, -1, 2) | 0.087 | 1.390 |
| (3, 1, 2) | 0.084 | 1.352 |
| (3, -1, 3) | 0.092 | 1.470 |
| (3, 1, 3) | 0.085 | 1.364 |
| (3, 2, 0) | 0.088 | 1.406 |
| (3, 2, 1) | 0.086 | 1.376 |
| (3, 2, 2) | 0.087 | 1.399 |
| (3, 2, 3) | 0.088 | 1.412 |
| (3, 3, 1) | 0.084 | 1.347 |
| (3, 3, 2) | 0.087 | 1.392 |

**ScZn**

| plane | γ [eV/Å$^2$] | γ [J/m$^2$] |
|---|---|---|
| (1, 0, 0) | 0.095 | 1.521 |
| (1, 1, 0) | 0.084 | 1.349 |
| (1, 1, 1) | 0.097 | 1.556 |
| (2, 1, 0) | 0.096 | 1.543 |
| (2, 1, 1) | 0.093 | 1.495 |
| (2, 2, 1) | 0.094 | 1.514 |
| (3, 1, 0) | 0.099 | 1.591 |
| (3, 1, 1) | 0.099 | 1.580 |
| (3, 2, 0) | 0.093 | 1.483 |
| (3, 2, 1) | 0.092 | 1.479 |
| (3, 2, 2) | 0.095 | 1.525 |
| (3, 3, 1) | 0.092 | 1.472 |
| (3, 3, 2) | 0.096 | 1.534 |

**$ScZn_2$**

| plane | γ [eV/Å$^2$] | γ [J/m$^2$] |
|---|---|---|
| (0, 0, 1) | 0.086 | 1.372 |
| (1, 0, 0) | 0.079 | 1.259 |
| (1, 0, 1) | 0.087 | 1.389 |
| (1, 0, 2) | 0.086 | 1.384 |
| (1, 0, 3) | 0.086 | 1.385 |
| (1, 1, 0) | 0.089 | 1.423 |
| (1, 1, 1) | 0.089 | 1.419 |
| (2, 0, 1) | 0.093 | 1.487 |
| (2, 0, 3) | 0.088 | 1.406 |
| (2, -1, 2) | 0.081 | 1.294 |
| (2, -1, 3) | 0.087 | 1.402 |
| (2, 2, 1) | 0.090 | 1.439 |
| (2, 2, 3) | 0.087 | 1.397 |
| (3, 0, 1) | 0.086 | 1.384 |
| (3, 0, 2) | 0.086 | 1.385 |
| (3, -1, 0) | 0.087 | 1.389 |
| (3, -1, 1) | 0.089 | 1.431 |
| (3, -1, 2) | 0.086 | 1.383 |
| (3, -1, 3) | 0.085 | 1.355 |
| (3, 1, 0) | 0.091 | 1.465 |
| (3, 1, 1) | 0.088 | 1.417 |
| (3, 1, 2) | 0.088 | 1.413 |
| (3, 1, 3) | 0.085 | 1.364 |
| (3, 2, 0) | 0.089 | 1.424 |
| (3, 2, 1) | 0.090 | 1.449 |
| (3, 2, 2) | 0.090 | 1.437 |
| (3, 2, 3) | 0.087 | 1.402 |
| (3, 3, 1) | 0.090 | 1.436 |
| (3, 3, 2) | 0.090 | 1.437 |

**Table S1.** Calculated surface energies of crystalline $Sc_xZn_y$ alloys from slab structures.

## $ScZn_3$

| plane | γ [eV/Å$^2$] | γ [J/m$^2$] |
|---|---|---|
| (0, 0, 1) | 0.070 | 1.127 |
| (1, 0, 0) | 0.078 | 1.244 |
| (1, 0, 1) | 0.079 | 1.259 |
| (1, 0, 2) | 0.087 | 1.397 |
| (1, 0, 3) | 0.083 | 1.337 |
| (1, -1, 0) | 0.067 | 1.071 |
| (1, 1, 0) | 0.083 | 1.331 |
| (1, -1, 1) | 0.090 | 1.450 |
| (1, 1, 1) | 0.084 | 1.349 |
| (1, -1, 2) | 0.087 | 1.398 |
| (1, 1, 2) | 0.089 | 1.434 |
| (1, -1, 3) | 0.083 | 1.338 |
| (1, 1, 3) | 0.089 | 1.421 |
| (2, 0, 1) | 0.097 | 1.556 |
| (2, 0, 3) | 0.087 | 1.387 |
| (2, -1, 0) | 0.082 | 1.316 |
| (2, 1, 0) | 0.083 | 1.333 |
| (2, -1, 1) | 0.088 | 1.413 |
| (2, 1, 1) | 0.087 | 1.392 |
| (2, -1, 2) | 0.090 | 1.443 |
| (2, 1, 2) | 0.087 | 1.392 |
| (2, -1, 3) | 0.090 | 1.441 |
| (2, 1, 3) | 0.089 | 1.426 |
| (2, -2, 1) | 0.075 | 1.203 |
| (2, 2, 1) | 0.087 | 1.388 |
| (2, -2, 3) | 0.087 | 1.396 |
| (2, 2, 3) | 0.089 | 1.418 |
| (3, 0, 1) | 0.076 | 1.225 |
| (3, 0, 2) | 0.080 | 1.285 |
| (3, -1, 0) | 0.083 | 1.332 |
| (3, 1, 0) | 0.088 | 1.408 |
| (3, -1, 1) | 0.085 | 1.357 |
| (3, 1, 1) | 0.084 | 1.339 |
| (3, -1, 2) | 0.087 | 1.394 |
| (3, 1, 2) | 0.083 | 1.323 |
| (3, -1, 3) | 0.090 | 1.436 |
| (3, 1, 3) | 0.087 | 1.401 |
| (3, -2, 0) | 0.084 | 1.342 |
| (3, 2, 0) | 0.083 | 1.323 |
| (3, -2, 1) | 0.086 | 1.371 |
| (3, 2, 1) | 0.087 | 1.391 |
| (3, -2, 2) | 0.088 | 1.416 |
| (3, 2, 2) | 0.087 | 1.390 |
| (3, -2, 3) | 0.088 | 1.417 |
| (3, 2, 3) | 0.086 | 1.375 |
| (3, -3, 1) | 0.075 | 1.195 |
| (3, 3, 1) | 0.087 | 1.389 |
| (3, -3, 2) | 0.086 | 1.375 |
| (3, 3, 2) | 0.087 | 1.392 |

## $Sc_{13}Zn_{73}$

| plane | γ [eV/Å$^2$] | γ [J/m$^2$] |
|---|---|---|
| (1, 0, 0) | 0.075 | 1.208 |
| (1, 1, 1) | 0.086 | 1.381 |
| (2, 1, 1) | 0.084 | 1.349 |
| (3, 1, 0) | 0.081 | 1.304 |
| (3, 2, 1) | 0.082 | 1.306 |
| (3, 3, 2) | 0.079 | 1.270 |
| (1, 1, 0) | 0.087 | 1.387 |
| (2, 1, 0) | 0.084 | 1.344 |
| (2, 2, 1) | 0.083 | 1.322 |
| (3, 1, 1) | 0.081 | 1.296 |
| (3, 2, 0) | 0.079 | 1.262 |
| (3, 2, 2) | 0.140 | 2.246 |
| (3, 3, 1) | 0.130 | 2.088 |

## $Sc_{13}Zn_{58}$

| plane | γ [eV/Å$^2$] | γ [J/m$^2$] |
|---|---|---|
| (0, 0, 1) | 0.088 | 1.403 |
| (1, 0, 0) | 0.088 | 1.407 |
| (1, 0, 1) | 0.090 | 1.442 |
| (1, 0, 2) | 0.080 | 1.287 |
| (1, 0, 3) | 0.081 | 1.291 |
| (1, 1, 0) | 0.092 | 1.471 |
| (1, 1, 1) | 0.082 | 1.310 |
| (2, -1, 2) | 0.084 | 1.347 |
| (2, -1, 3) | 0.080 | 1.276 |
| (2, 0, 1) | 0.084 | 1.340 |
| (2, 0, 3) | 0.082 | 1.315 |
| (2, 2, 1) | 0.077 | 1.236 |
| (2, 2, 3) | 0.075 | 1.202 |
| (3, -1, 0) | 0.086 | 1.380 |
| (3, -1, 1) | 0.076 | 1.211 |
| (3, -1, 2) | 0.080 | 1.285 |
| (3, -1, 3) | 0.077 | 1.230 |
| (3, 0, 1) | 0.082 | 1.311 |
| (3, 0, 2) | 0.080 | 1.285 |
| (3, 1, 0) | 0.078 | 1.252 |
| (3, 1, 1) | 0.079 | 1.266 |
| (3, 1, 2) | 0.080 | 1.285 |
| (3, 1, 3) | 0.075 | 1.199 |
| (3, 2, 0) | 0.072 | 1.159 |
| (3, 2, 1) | 0.075 | 1.205 |
| (3, 2, 2) | 0.074 | 1.181 |
| (3, 2, 3) | 0.078 | 1.242 |
| (3, 3, 1) | 0.072 | 1.155 |
| (3, 3, 2) | 0.071 | 1.145 |

**Table S1.** Continued.

**$ScZn_6$**

| plane | γ [eV/Å$^2$] | γ [J/m$^2$] |
|---|---|---|
| (1, 0, 0) | 0.083 | 1.335 |
| (1, 1, 1) | 0.087 | 1.397 |
| (2, 1, 1) | 0.083 | 1.324 |
| (3, 1, 0) | 0.083 | 1.333 |
| (3, 2, 1) | 0.084 | 1.346 |
| (3, 3, 2) | 0.082 | 1.319 |
| (1, 1, 0) | 0.078 | 1.256 |
| (2, 1, 0) | 0.087 | 1.398 |
| (2, 2, 1) | 0.084 | 1.353 |
| (3, 1, 1) | 0.077 | 1.236 |
| (3, 2, 0) | 0.084 | 1.353 |
| (3, 2, 2) | 0.104 | 1.669 |
| (3, 3, 1) | 0.083 | 1.330 |

**$ScZn_{12}$**

| plane | γ [eV/Å$^2$] | γ [J/m$^2$] |
|---|---|---|
| (0, 0, 1) | 0.070 | 1.126 |
| (1, 0, 0) | 0.080 | 1.277 |
| (1, 0, 1) | 0.071 | 1.139 |
| (1, 0, 2) | 0.069 | 1.110 |
| (1, 0, 3) | 0.074 | 1.188 |
| (1, 1, 0) | 0.069 | 1.112 |
| (1, 1, 1) | 0.082 | 1.316 |
| (1, 1, 2) | 0.073 | 1.175 |
| (1, 1, 3) | 0.072 | 1.154 |
| (2, 0, 1) | 0.076 | 1.222 |
| (2, 0, 3) | 0.070 | 1.129 |
| (2, 1, 0) | 0.079 | 1.260 |
| (2, 1, 1) | 0.073 | 1.172 |
| (2, 1, 2) | 0.074 | 1.179 |
| (2, 1, 3) | 0.068 | 1.085 |
| (2, 2, 1) | 0.074 | 1.183 |
| (2, 2, 3) | 0.069 | 1.102 |
| (3, 0, 1) | 0.073 | 1.174 |
| (3, 0, 2) | 0.070 | 1.121 |
| (3, 1, 0) | 0.071 | 1.140 |
| (3, 1, 1) | 0.075 | 1.195 |
| (3, 1, 2) | 0.072 | 1.146 |
| (3, 1, 3) | 0.070 | 1.122 |
| (3, 2, 0) | 0.074 | 1.187 |
| (3, 2, 1) | 0.071 | 1.142 |
| (3, 2, 2) | 0.072 | 1.161 |
| (3, 2, 3) | 0.067 | 1.080 |
| (3, 3, 1) | 0.075 | 1.197 |
| (3, 3, 2) | 0.071 | 1.134 |

**Zn**

| plane | γ [eV/Å$^2$] | γ [J/m$^2$] |
|---|---|---|
| (0, 0, 1) | 0.070 | 1.126 |
| (1, 0, 0) | 0.080 | 1.277 |
| (1, 0, 1) | 0.071 | 1.139 |
| (1, 0, 2) | 0.069 | 1.110 |
| (1, 0, 3) | 0.074 | 1.188 |
| (1, 1, 0) | 0.069 | 1.112 |
| (1, 1, 1) | 0.082 | 1.316 |
| (1, 1, 2) | 0.073 | 1.175 |
| (1, 1, 3) | 0.072 | 1.154 |
| (2, 0, 1) | 0.076 | 1.222 |
| (2, 0, 3) | 0.070 | 1.129 |
| (2, 1, 0) | 0.079 | 1.260 |
| (2, 1, 1) | 0.073 | 1.172 |
| (2, 1, 2) | 0.074 | 1.179 |
| (2, 1, 3) | 0.068 | 1.085 |
| (2, 2, 1) | 0.074 | 1.183 |
| (2, 2, 3) | 0.069 | 1.102 |
| (3, 0, 1) | 0.073 | 1.174 |
| (3, 0, 2) | 0.070 | 1.121 |
| (3, 1, 0) | 0.071 | 1.140 |
| (3, 1, 1) | 0.075 | 1.195 |
| (3, 1, 2) | 0.072 | 1.146 |
| (3, 1, 3) | 0.070 | 1.122 |
| (3, 2, 0) | 0.074 | 1.187 |
| (3, 2, 1) | 0.071 | 1.142 |
| (3, 2, 2) | 0.072 | 1.161 |
| (3, 2, 3) | 0.067 | 1.080 |
| (3, 3, 1) | 0.075 | 1.197 |
| (3, 3, 2) | 0.071 | 1.134 |

**Table S1.** Continued.

**Yb**

| plane | γ [eV/Å$^2$] | γ [J/m$^2$] |
|---|---|---|
| (0, 0, 1) | 0.031 | 0.500 |
| (1, 0, 0) | 0.035 | 0.563 |
| (1, 0, 1) | 0.033 | 0.526 |
| (1, 0, 2) | 0.033 | 0.528 |
| (1, 0, 3) | 0.033 | 0.522 |
| (1, 1, 0) | 0.034 | 0.553 |
| (1, 1, 1) | 0.035 | 0.566 |
| (2, 0, 1) | 0.036 | 0.570 |
| (2, 0, 3) | 0.034 | 0.551 |
| (2, -1, 2) | 0.036 | 0.573 |
| (2, -1, 3) | 0.037 | 0.589 |
| (2, 2, 1) | 0.035 | 0.560 |
| (2, 2, 3) | 0.036 | 0.572 |
| (3, 0, 1) | 0.035 | 0.568 |
| (3, 0, 2) | 0.035 | 0.555 |
| (3, -1, 0) | 0.036 | 0.573 |
| (3, 1, 0) | 0.036 | 0.579 |
| (3, -1, 1) | 0.035 | 0.562 |
| (3, 1, 1) | 0.041 | 0.660 |
| (3, -1, 2) | 0.035 | 0.566 |
| (3, 1, 2) | 0.041 | 0.658 |
| (3, -1, 3) | 0.035 | 0.569 |
| (3, 1, 3) | 0.036 | 0.574 |
| (3, 2, 0) | 0.035 | 0.568 |
| (3, 2, 1) | 0.039 | 0.631 |
| (3, 2, 2) | 0.040 | 0.638 |
| (3, 2, 3) | 0.035 | 0.566 |
| (3, 3, 1) | 0.035 | 0.556 |
| (3, 3, 2) | 0.035 | 0.561 |

**YbCd**

| plane | γ [eV/Å$^2$] | γ [J/m$^2$] |
|---|---|---|
| (1, 0, 0) | 0.045 | 0.726 |
| (1, 1, 0) | 0.039 | 0.632 |
| (1, 1, 1) | 0.045 | 0.713 |
| (2, 1, 0) | 0.045 | 0.723 |
| (2, 1, 1) | 0.043 | 0.695 |
| (2, 2, 1) | 0.044 | 0.713 |
| (3, 1, 0) | 0.047 | 0.755 |
| (3, 1, 1) | 0.046 | 0.740 |
| (3, 2, 0) | 0.044 | 0.703 |
| (3, 2, 1) | 0.043 | 0.693 |
| (3, 2, 2) | 0.045 | 0.719 |
| (3, 3, 1) | 0.044 | 0.701 |
| (3, 3, 2) | 0.045 | 0.718 |

**$YbCd_2$**

| plane | γ [eV/Å$^2$] | γ [J/m$^2$] |
|---|---|---|
| (0, 0, 1) | 0.045 | 0.713 |
| (1, 0, 0) | 0.041 | 0.650 |
| (1, 0, 1) | 0.045 | 0.716 |
| (1, 0, 2) | 0.044 | 0.697 |
| (1, 0, 3) | 0.048 | 0.774 |
| (1, 1, 0) | 0.046 | 0.730 |
| (1, 1, 1) | 0.046 | 0.730 |
| (2, -1, 2) | 0.042 | 0.674 |
| (2, -1, 3) | 0.045 | 0.723 |
| (2, 0, 1) | 0.043 | 0.689 |
| (2, 0, 3) | 0.046 | 0.733 |
| (2, 2, 1) | 0.046 | 0.739 |
| (2, 2, 3) | 0.045 | 0.718 |
| (3, -1, 0) | 0.045 | 0.716 |
| (3, -1, 1) | 0.046 | 0.738 |
| (3, -1, 2) | 0.045 | 0.716 |
| (3, -1, 3) | 0.045 | 0.721 |
| (3, 0, 1) | 0.043 | 0.683 |
| (3, 0, 2) | 0.044 | 0.709 |
| (3, 1, 0) | 0.047 | 0.756 |
| (3, 1, 1) | 0.046 | 0.736 |
| (3, 1, 2) | 0.046 | 0.732 |
| (3, 1, 3) | 0.045 | 0.713 |
| (3, 2, 0) | 0.046 | 0.744 |
| (3, 2, 1) | 0.047 | 0.752 |
| (3, 2, 2) | 0.046 | 0.733 |
| (3, 2, 3) | 0.046 | 0.735 |
| (3, 3, 1) | 0.046 | 0.738 |
| (3, 3, 2) | 0.046 | 0.739 |

**Table S2.** Calculated surface energies of crystalline $Yb_xCd_y$ alloys from slab structures.

## $YbCd_3$

| plane | γ [eV/Å$^2$] | γ [J/m$^2$] |
|---|---|---|
| (0, 0, 1) | 0.037 | 0.596 |
| (1, 0, 0) | 0.038 | 0.603 |
| (1, 0, 1) | 0.040 | 0.643 |
| (1, 0, 2) | 0.043 | 0.695 |
| (1, 0, 3) | 0.042 | 0.670 |
| (1, 1, 0) | 0.041 | 0.650 |
| (1, 1, 1) | 0.041 | 0.655 |
| (2, 0, 1) | 0.047 | 0.754 |
| (2, 0, 3) | 0.043 | 0.690 |
| (2, -1, 2) | 0.044 | 0.699 |
| (2, -1, 3) | 0.044 | 0.703 |
| (2, 2, 1) | 0.043 | 0.684 |
| (2, 2, 3) | 0.044 | 0.698 |
| (3, 0, 1) | 0.038 | 0.606 |
| (3, 0, 2) | 0.039 | 0.632 |
| (3, -1, 0) | 0.041 | 0.660 |
| (3, 1, 0) | 0.043 | 0.684 |
| (3, -1, 1) | 0.042 | 0.665 |
| (3, 1, 1) | 0.042 | 0.674 |
| (3, -1, 2) | 0.043 | 0.694 |
| (3, 1, 2) | 0.040 | 0.639 |
| (3, -1, 3) | 0.044 | 0.701 |
| (3, 1, 3) | 0.043 | 0.686 |
| (3, 2, 0) | 0.041 | 0.650 |
| (3, 2, 1) | 0.042 | 0.679 |
| (3, 2, 2) | 0.042 | 0.671 |
| (3, 2, 3) | 0.042 | 0.680 |
| (3, 3, 1) | 0.042 | 0.681 |
| (3, 3, 2) | 0.042 | 0.680 |

## $Yb_{14}Cd_{51}$

| plane | γ [eV/Å$^2$] | γ [J/m$^2$] |
|---|---|---|
| (0, 0, 1) | 0.041 | 0.657 |
| (1, 0, 0) | 0.040 | 0.641 |
| (1, 0, 1) | 0.044 | 0.704 |
| (1, 0, 2) | 0.040 | 0.633 |
| (1, 0, 3) | 0.037 | 0.591 |
| (1, 1, 0) | 0.041 | 0.659 |
| (1, 1, 1) | 0.041 | 0.652 |
| (2, -1, 2) | 0.039 | 0.629 |
| (2, -1, 3) | 0.039 | 0.629 |
| (2, 0, 1) | 0.041 | 0.653 |
| (2, 0, 3) | 0.037 | 0.591 |
| (2, 2, 1) | 0.039 | 0.621 |
| (2, 2, 3) | 0.036 | 0.574 |
| (3, -1, 0) | 0.039 | 0.621 |
| (3, -1, 1) | 0.041 | 0.654 |
| (3, -1, 2) | 0.038 | 0.615 |
| (3, -1, 3) | 0.036 | 0.569 |
| (3, 0, 1) | 0.040 | 0.639 |
| (3, 0, 2) | 0.037 | 0.596 |
| (3, 1, 0) | 0.039 | 0.630 |
| (3, 1, 1) | 0.039 | 0.618 |
| (3, 1, 2) | 0.034 | 0.549 |
| (3, 1, 3) | 0.034 | 0.549 |
| (3, 2, 0) | 0.036 | 0.579 |
| (3, 2, 1) | 0.037 | 0.593 |
| (3, 2, 2) | 0.036 | 0.571 |
| (3, 2, 3) | 0.034 | 0.548 |
| (3, 3, 1) | 0.034 | 0.549 |
| (3, 3, 2) | 0.034 | 0.549 |

## $Yb_{13}Cd_{76}$

| plane | γ [eV/Å$^2$] | γ [J/m$^2$] |
|---|---|---|
| (1, 0, 0) | 0.041 | 0.656 |
| (1, 1, 1) | 0.697 | 11.168 |
| (2, 1, 1) | 0.045 | 0.719 |
| (3, 1, 0) | 0.043 | 0.692 |
| (3, 2, 1) | 0.037 | 0.599 |
| (3, 3, 2) | 0.041 | 0.658 |
| (1, 1, 0) | 0.045 | 0.726 |
| (2, 1, 0) | 0.043 | 0.696 |
| (2, 2, 1) | 0.154 | 2.474 |
| (3, 1, 1) | 0.123 | 1.966 |
| (3, 2, 0) | 0.041 | 0.652 |
| (3, 2, 2) | 0.084 | 1.340 |
| (3, 3, 1) | 0.075 | 1.208 |

**Table S2.** Continued.

## $YbCd_6$

| plane | γ [eV/Å$^2$] | γ [J/m$^2$] |
|---|---|---|
| (1, 0, 0) | 0.045 | 0.717 |
| (1, 1, 0) | 0.044 | 0.701 |
| (1, 1, 1) | 0.045 | 0.720 |
| (2, 1, 1) | 0.045 | 0.729 |
| (3, 1, 0) | 0.045 | 0.723 |
| (3, 2, 1) | 0.046 | 0.736 |
| (3, 3, 2) | 0.045 | 0.723 |
| (2, 1, 0) | 0.048 | 0.766 |
| (2, 2, 1) | 0.046 | 0.731 |
| (3, 1, 1) | 0.060 | 0.969 |
| (3, 2, 0) | 0.046 | 0.735 |
| (3, 2, 2) | 0.057 | 0.920 |
| (3, 3, 1) | 0.046 | 0.730 |

## Cd

| plane | γ [eV/Å$^2$] | γ [J/m$^2$] |
|---|---|---|
| (0, 0, 1) | 0.017 | 0.277 |
| (1, 0, 0) | 0.042 | 0.666 |
| (1, 0, 1) | 0.036 | 0.584 |
| (1, 0, 2) | 0.036 | 0.572 |
| (1, 0, 3) | 0.031 | 0.491 |
| (1, 1, 0) | 0.040 | 0.637 |
| (1, 1, 1) | 0.040 | 0.639 |
| (2, 0, 1) | 0.038 | 0.614 |
| (2, 0, 3) | 0.033 | 0.534 |
| (2, -1, 2) | 0.038 | 0.610 |
| (2, -1, 3) | 0.038 | 0.604 |
| (2, 2, 1) | 0.039 | 0.629 |
| (2, 2, 3) | 0.038 | 0.617 |
| (3, 0, 1) | 0.036 | 0.577 |
| (3, 0, 2) | 0.034 | 0.538 |
| (3, -1, 0) | 0.042 | 0.670 |
| (3, 1, 0) | 0.036 | 0.584 |
| (3, -1, 1) | 0.041 | 0.653 |
| (3, 1, 1) | 0.039 | 0.627 |
| (3, -1, 2) | 0.040 | 0.635 |
| (3, 1, 2) | 0.037 | 0.594 |
| (3, -1, 3) | 0.038 | 0.613 |
| (3, 1, 3) | 0.037 | 0.600 |
| (3, 2, 0) | 0.041 | 0.657 |
| (3, 2, 1) | 0.040 | 0.641 |
| (3, 2, 2) | 0.040 | 0.641 |
| (3, 2, 3) | 0.040 | 0.633 |
| (3, 3, 1) | 0.039 | 0.626 |
| (3, 3, 2) | 0.039 | 0.630 |

**Table S2.** Continued.

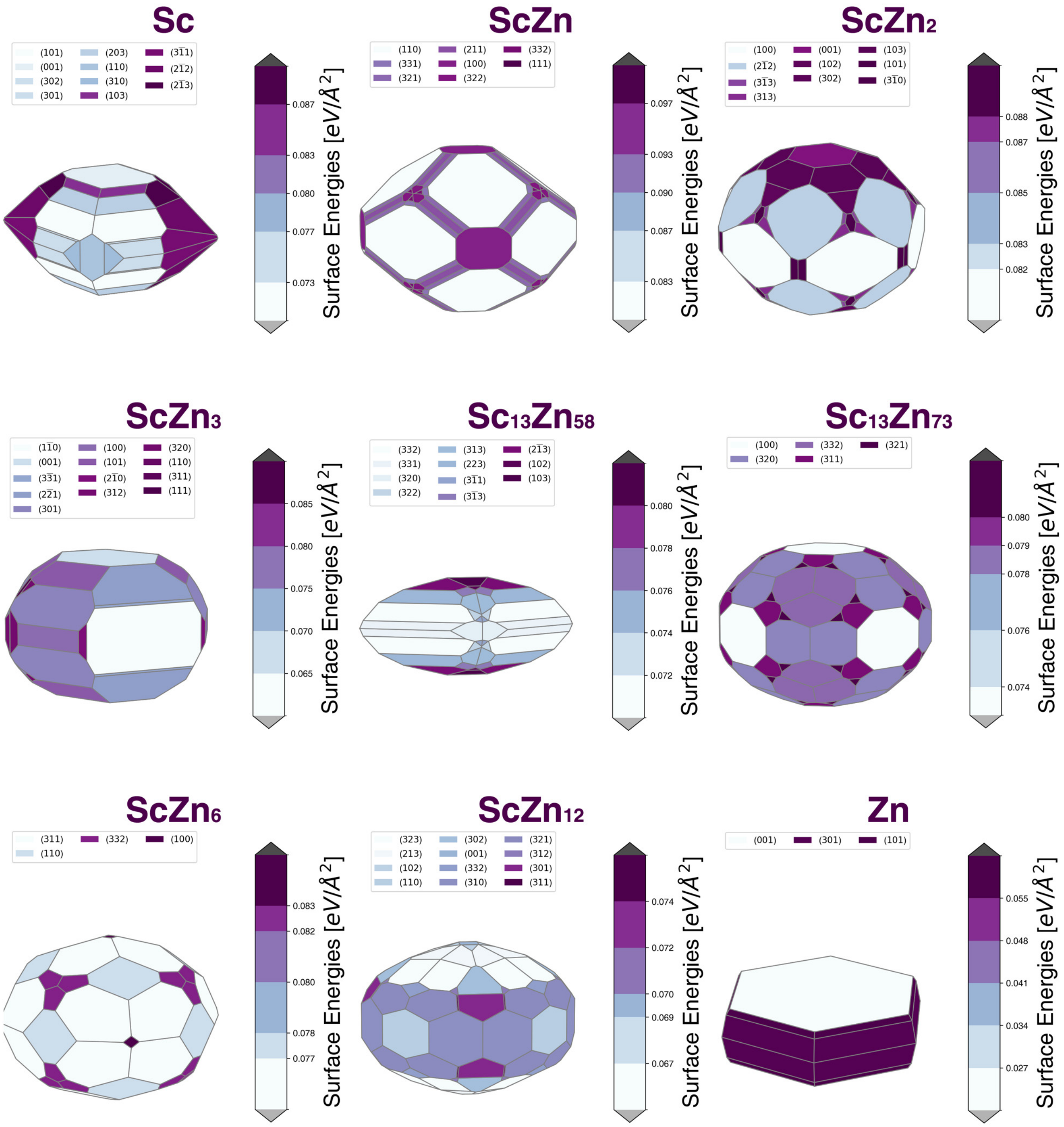

**Figure S9.** Calculated Wulff shapes of crystalline $Sc_xZn_y$ alloys.

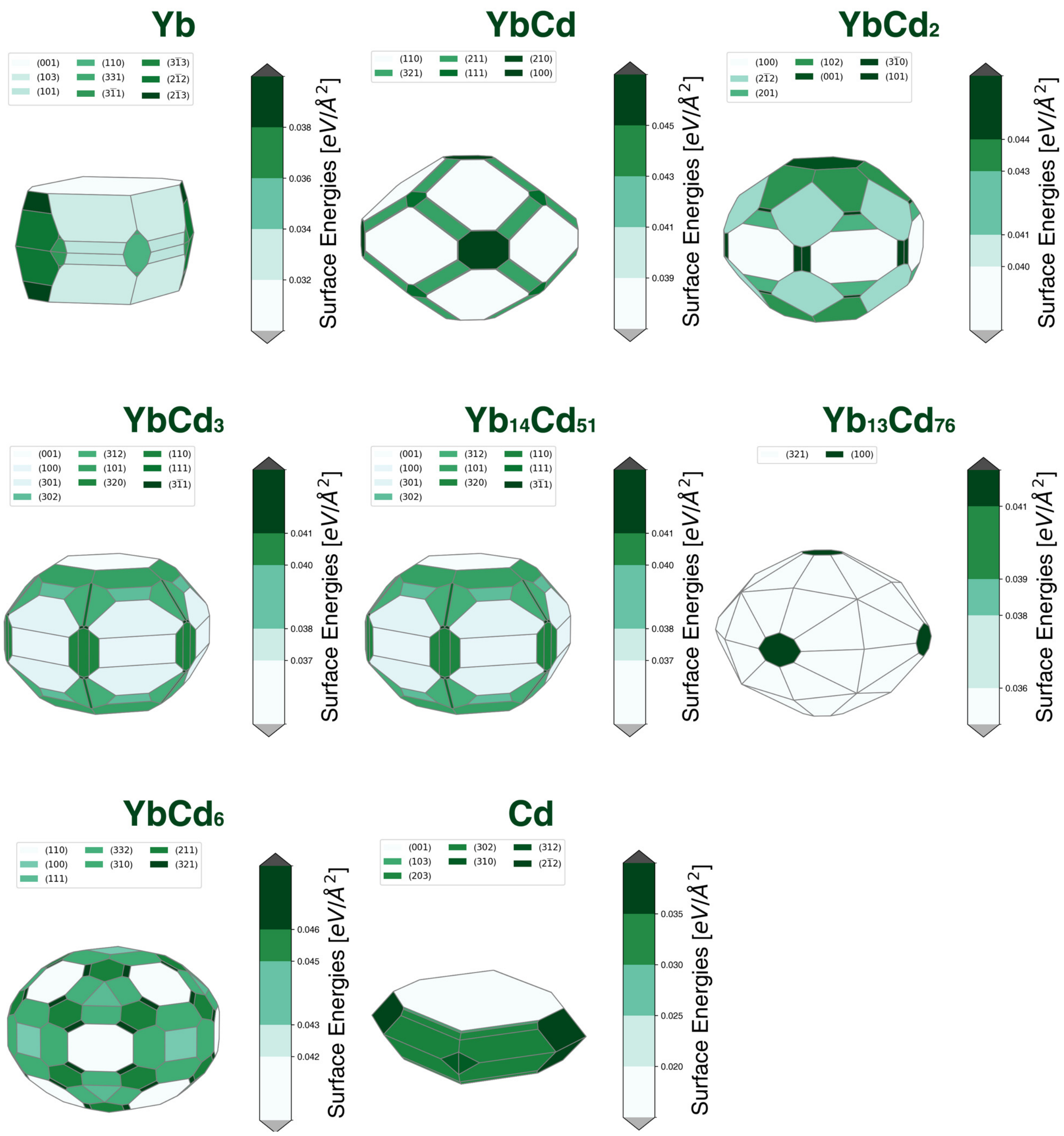

**Figure S10.** Calculated Wulff shapes of crystalline $Yb_xCd_y$ alloys.

| | Sc | ScZn | $ScZn_2$ | $ScZn_3$ | $Sc_{13}Zn_{58}$ | $Sc_{13}Zn_{73}$ | $ScZn_6$ | iQC-$ScZn_{7.33}$ | $ScZn_{12}$ | Zn |
|---|---|---|---|---|---|---|---|---|---|---|
| Shape factor (η) | 5.050 | 5.060 | 4.994 | 5.098 | 5.533 | 4.924 | 4.959 | 4.836 | 4.931 | 6.017 |
| Atomic density (ρ) [atom/Å$^3$] | 0.040 | 0.053 | 0.059 | 0.060 | 0.062 | 0.062 | 0.063 | 0.064 | 0.065 | 0.064 |
| Weighted surface energy ($\gamma$) [eV/Å$^2$] | 0.082 | 0.088 | 0.082 | 0.075 | 0.074 | 0.079 | 0.078 | 0.076 | 0.070 | 0.041 |
| Weighted surface energy ($\gamma$) [J/m$^2$] | 1.306 | 1.402 | 1.313 | 1.196 | 1.189 | 1.258 | 1.249 | 1.224 | 1.119 | 0.660 |
| Bulk formation energy ($E_{bulk}$/N) [eV/atom] | 0 | -0.382 | -0.350 | -0.312 | -0.261 | -0.241 | -0.237 | -0.222 | -0.130 | 0 |

**Table S3.** Calculated parameters for the Wulff constructions of crystalline $Sc_xZn_y$ systems from the periodic and slab calculations. The parameters of iQC $ScZn_{7.33}$ are calculated from the finite-size DFT calculations.

| | Yb | YbCd | $YbCd_2$ | $YbCd_3$ | $Yb_{14}Cd_{51}$ | $Yb_{13}Cd_{76}$ | iQC-$YbCd_{5.7}$ | $YbCd_6$ | Cd |
|---|---|---|---|---|---|---|---|---|---|
| Shape factor (η) | 5.310 | 5.063 | 4.999 | 4.978 | 4.950 | 4.966 | 4.836 | 4.903 | 5.632 |
| Atomic density (ρ) [atom/Å$^3$] | 0.026 | 0.036 | 0.039 | 0.040 | 0.041 | 0.042 | 0.043 | 0.042 | 0.043 |
| Weighted surface energy ($\gamma$) [eV/Å$^2$] | 0.033 | 0.041 | 0.042 | 0.039 | 0.035 | 0.038 | 0.040 | 0.045 | 0.028 |
| Weighted surface energy ($\gamma$) [J/m$^2$] | 0.532 | 0.657 | 0.679 | 0.621 | 0.554 | 0.602 | 0.633 | 0.718 | 0.446 |
| Bulk formation energy ($E_{bulk}$/N) [eV/atom] | 0 | -0.334 | -0.298 | -0.253 | -0.249 | -0.180 | -0.201 | -0.188 | 0 |

**Table S4.** Calculated parameters for the Wulff constructions of crystalline $Yb_xCd_y$ systems from the periodic and slab calculations. The parameters of iQC $YbCd_{5.7}$ are calculated from the finite-size DFT calculations.

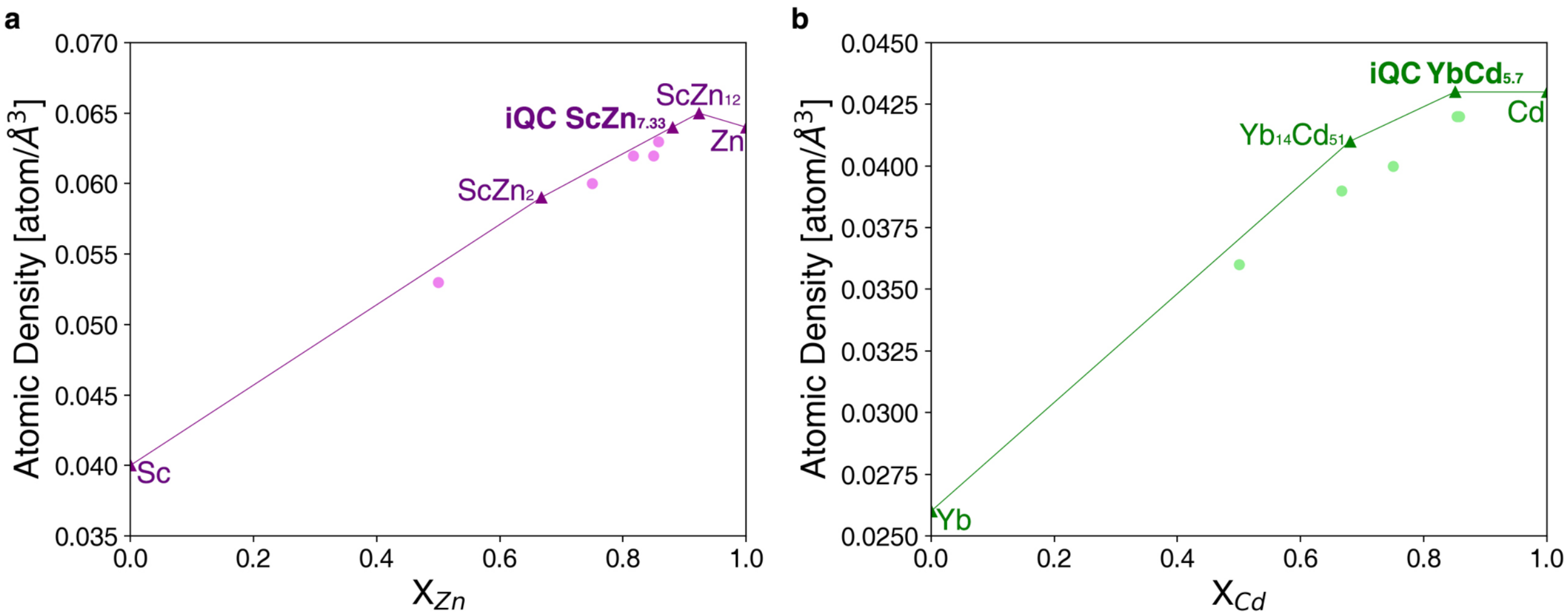

**Figure S11.** Calculated hull of atomic density of **a,** $Sc_xZn_y$ and **b,** $Yb_xCd_y$. The lines connecting triangular points show the densest combination of alloys for a given composition. The quasicrystals represent the densest possible intermetallic arrangements in their chemical system at their composition.

## S6. Calculation of Phase Diagram

### Prior thermodynamic assessments

Tang et al. and Kali-Ali et al. have previously reported thermodynamic modeling of the Sc-Zn and Yb-Cd systems, aided by *ab initio* calculations.[21,22] The thermodynamic models of these two systems are reassessed here using the CALPHAD approach, incorporating our DFT formation energies of all structures including the quasicrystals.

Palenzona et al. reported phase diagram data for the Sc-Zn system.[23] They measured phase equilibria in the range of 40 to 100 % $X_{Zn}$ using differential thermal analysis (DTA), X-ray diffraction (XRD), and electron microscopy. Five intermetallic compounds were identified in this system: ScZn, $ScZn_2$, $Sc_{17}Zn_{58}$, $Sc_3Zn_{17}$, and $ScZn_{12}$. However, further research into the Sc-Cu-Zn system led to a proposed correction of stoichiometry from $Sc_3Zn_{17}$ to $ScZn_6$.[24] This correction was prompted by the discovery of the $Sc_3Cu_yZn_{18-y}$ phase region during explorations of $Sc_3Cu_xZn_{17-x}$ compositions, where copper substitution for zinc suggested a more stable and representative $YCd_6$ ($RCd_6$) type structure, akin to that of $ScZn_6$. Additionally, Canfield et al. reported a binary icosahedral quasicrystal phase, $Sc_{12}Zn_{88}$, which decomposes through a peritectic reaction at 778 K.[25] The experimental thermodynamic data for the Sc-Zn system remain absent.

Palenzona determined the experimental Yb-Cd phase diagram data through DTA, XRD, and electron microscopy.[26] Six intermetallic compounds were identified in this system: YbCd, $YbCd_2$, $Yb_3Cd_8$, $Yb_{14}Cd_{51}$, $YbCd_{5.7}$ and $YbCd_6$. A polymorphic transition of $YbCd_2$ was detected at 968 K ($\alpha$YbCd2 → $\beta$YbCd2). The crystal structure of some compounds was not experimentally defined. Later, $YbCd_{5.7}$ was found to be a binary icosahedral quasicrystal phase with a congruent melting point at 909 K.[18] Currently, there is no experimental thermodynamic data available for the Yb-Cd system.

### Thermodynamic Models

1. Unary phases

For the pure element of i (i = Cd, Sc, Yb, Zn) in φ phase (φ = liquid, Cd (hcp), Sc (bcc), Sc (hcp), Yb (fcc), Yb (bcc), or Zn (hcp)), the Gibbs free energy function ${}^0G_i^\varphi = G_i^\varphi - H_i^{SER}$ of temperature is given as follows:

$${}^0G_i^\varphi = a + bT + cT\ln T + dT^2 + eT^3 + fT^{-1} + gT^7 + hT^{-9}$$

where $H_i^{SER}$ is the molar enthalpy of the element *i* at 298.15 K and 1 bar in its stable element reference (SER) state, and T is the absolute temperature. The coefficients are taken from the Scientific Group Thermodata Europe (SGTE) database compilation by Dinsdale.[27]

2. Solution phases

For the solution phases φ (φ = liquid, Yb (fcc), Yb (bcc)), the Gibbs free energy function is given as follows:

$$G^\varphi = x_i{}^0G_i^\varphi + x_j{}^0G_i^\varphi + RT(x_i\ln x_i + x_j\ln x_j) + {}^E G^\varphi$$

where $x_i$ and $x_j$ are the mole fraction of the components i and j, R is the gas constant, and ${}^E G^\varphi$ is the excess Gibbs energy which can be described by Relich-Kister binary excess model[28]

$$^{E}G^{\varphi} = x_i x_j \sum_{v=0}^{k} (x_i - x_j)^v \cdot {}^v L_{ij}$$

$$^{v}L_{ij} = {}^{v}a_{ij} + {}^{v}b_{ij}T$$

where ${}^{v}L_{ij}$ represents the temperature-dependent interaction parameters, and ${}^{v}a_{ij}$ and ${}^{v}b_{ij}$ are the parameters to be evaluated from numerical optimization.

3. Stochiometric phases

For the stochiometric phases φ (φ = ScZn, $ScZn_2$, $Sc_{13}Zn_{58}$, $ScZn_6$, $ScZn_{12}$, YbCd, $\alpha YbCd_2$, $\beta YbCd_2$, $Yb_3Cd_8$, $Yb_{14}Cd_{51}$, $YbCd_{5.7}$, and $YbCd_6$), the Gibbs free energy function is given as follows:

$$G^{\varphi} = x_i {}^0G_i^{\varphi} + x_j {}^0G_i^{\varphi} + \Delta_f H^{\varphi} - T\Delta_f S^{\varphi}$$

where $\Delta_f H^{\varphi}$ is the enthalpy of formation of the stochiometric phase φ and $\Delta_f S^{\varphi}$ is the entropy of formation of the stochiometric phase φ. Most enthalpy of formation data is obtained through DFT calculation, except for $\beta YbCd_2$, $Yb_3Cd_8$ due to lack of structure information.

**Optimization procedures**

The thermodynamic parameters in the Sc-Zn and Yb-Cd systems were initially generated and optimized using the Extensible Self-optimized Phase Equilibria Infrastructure (ESPEI) package.[29] In this process, ESPEI first generates model parameters for CALPHAD models of the Gibbs energy for end members and intermetallic compounds, and Redlich-Kister interaction parameters based on the input of thermochemistry data using Muggianu extrapolation.[30] Due to the lack of experimental data on liquid mixing energies, the initial interaction *L* parameters were derived by calculating the differences in Gibbs energies of the liquids and solids at various phase boundaries. For instance, at congruent melting points, the enthalpy of non-ideal mixing in the liquid phase can be estimated given that the ideal liquid and solid energies (with solid entropy sourced from the literature) are known.

Interaction parameters and formation entropies of intermetallic compounds were simultaneously optimized using ESPEI via the Bayesian ensemble Markov Chain Monte Carlo (MCMC) method[31] employing triangular priors ranging from ± 0.8 θ where θ represents the initial parameter. Utilizing only the estimated liquid mixing energy and the optimized formation entropy of intermetallic compounds from literature, the optimized thermodynamic parameters in both systems were unable to perfectly fit the liquidus portion after 500 MCMC iterations. Consequently, a second assessment was conducted. This stage involved a meticulous adjustment of the formation entropy values for intermetallic compounds. Adjustments were made incrementally, and the phase diagram was recalculated after each modification to calibrate the numerical deviations. This iterative process continued until the calculated phase diagram aligned well with the experimental data within the expected uncertainty limits. The resulting phase diagrams of the Sc-Zn and Yb-Cd systems are shown in **Figure S12**, and optimized parameters of liquids and solids are summarized in **Table S5 and S6**.

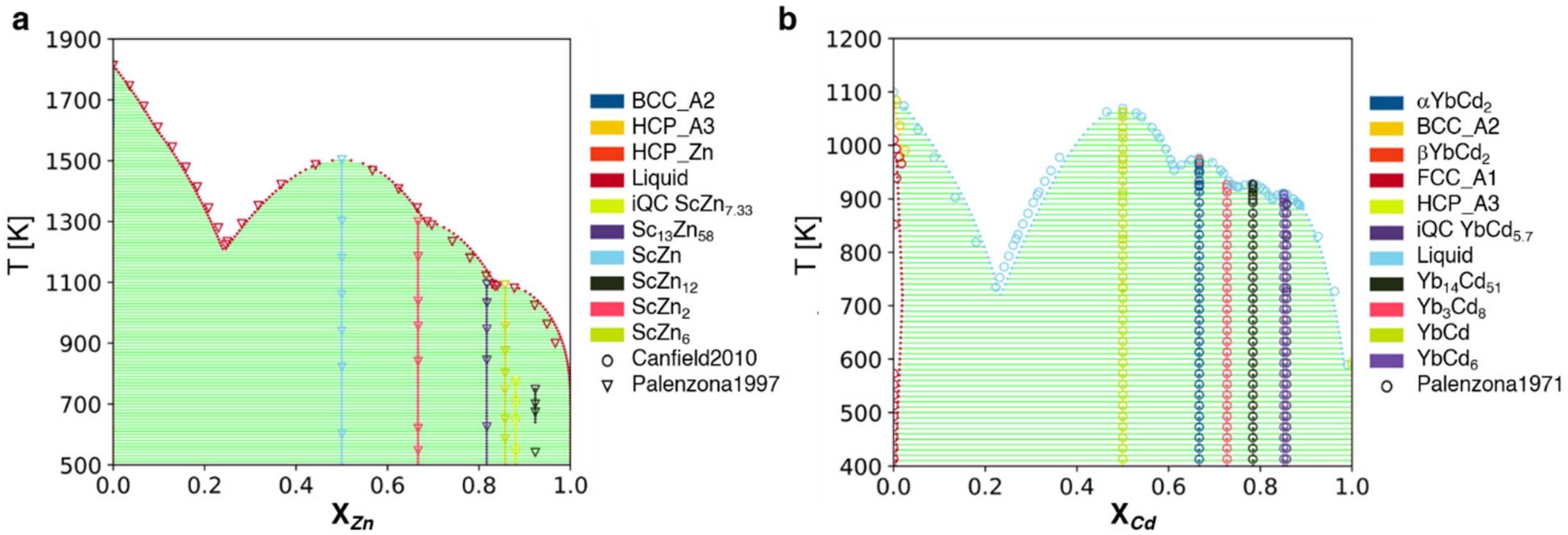

**Figure S12.** Optimized **a,** Sc-Zn and **b,** Yb-Cd phase diagrams using ESPEI (dots and lines) and the experimental phase equilibrium data (denoted by triangles and circles) from Materials Platform for Data Science (MPDS).[32]

| | Parameters |
|---|---|
| Liquid | $L_0 = -83967 + 10.7848$ T,<br>$L_1 = 15273$ |
| ScZn | $\Delta_f H^\varphi = -36859$, $\Delta_f S^\varphi = -5.4856$ |
| $ScZn_2$ | $\Delta_f H^\varphi = -33772$, $\Delta_f S^\varphi = -5.7182$ |
| $Sc_{13}Zn_{58}$ | $\Delta_f H^\varphi = -25143$, $\Delta_f S^\varphi = -5.6088$ |
| $ScZn_6$ | $\Delta_f H^\varphi = -22841$, $\Delta_f S^\varphi = -5.7117$ |
| iQC-$ScZn_{7.33}$ | $\Delta_f H^\varphi = -21467$, $\Delta_f S^\varphi = -7.5554$ |
| $ScZn_{12}$ | $\Delta_f H^\varphi = -13583$, $\Delta_f S^\varphi = -4.3402$ |

**Table S5.** Summary of optimized thermodynamic parameters of alloys in the Sc-Zn system from the ESPEI.

| | Parameters |
|---|---|
| Liquid | $L_0 = -94745 + 3.7720$ T,<br>$L_1 = -31759 + 0.5764$ T |
| YbCd | $\Delta_f H^\varphi = -32253$, $\Delta_f S^\varphi = -1.0098$ |
| $\alpha YbCd_2$ | $\Delta_f H^\varphi = -28764$, $\Delta_f S^\varphi = 1.3293$ |
| $\beta YbCd_2$ | $\Delta_f H^\varphi = -27932$, $\Delta_f S^\varphi = 2.1884$ |
| $Yb_3Cd_8$ | $\Delta_f H^\varphi = -27965$, $\Delta_f S^\varphi = -0.4836$ |
| $Yb_{14}Cd_{51}$ | $\Delta_f H^\varphi = -24064$, $\Delta_f S^\varphi = 0.8416$ |
| iQC-$YbCd_{5.7}$ | $\Delta_f H^\varphi = -19355$, $\Delta_f S^\varphi = 0.8317$ |
| $YbCd_6$ | $\Delta_f H^\varphi = -18603$, $\Delta_f S^\varphi = 0.9936$ |

**Table S6.** Summary of optimized thermodynamic parameters of alloys in the Yb-Cd system from the ESPEI.

In the calculated Sc-Zn system, as shown in **Figure S12a**, it should be noted that the $ScZn_{12}$ phase exhibits metastability at the ground state (0 K), presenting a discrepancy with the previously published phase diagrams. Moreover, $Sc_{13}Zn_{58}$ and $ScZn_6$ phases are also thermodynamically metastable at the ground state, but this detail is not evident due to the reported phase diagram's lowest temperature threshold of 500 K. This metastability is attributed to the formation energies of these phases, as calculated by DFT, which do not lie on the convex hull, thereby making them metastable at low temperature. This discrepancy may indeed reflect actual phase behaviors, as the sample used in DTA was prepared via the cooling of a mixed liquid during the experimental measurement of the Sc-Zn phase diagram. Throughout this process, kinetic barriers may inhibit the system from attaining its thermodynamic ground state. Notably, if the energy barrier required for the transformation from a metastable to a stable phase is substantial, the metastable phases ($ScZn_6$ and $Sc_{13}Zn_{58}$) can persist after solidification. Consequently, a phase might be routinely observed in such method of measuring phase equilibrium data, even if it is not the most stable thermodynamically, due to the system's insufficient energy or time to overcome the barrier and transition into a more stable phase.

## S7. Size-dependent convex hull and nucleation analysis of the Yb-Cd system

From the calculated parameters, theoretical total formation energy of a nanocluster of crystalline alloys can be calculated as a function of Area-to-Volume ratio, or equivalently, $1/R$ (1/R).[33]

$$\frac{E_{NP}}{N} = \frac{E_{bulk}}{N} + \frac{\gamma\eta}{\rho} \cdot \frac{1}{R}$$

where $\eta$ is dimensionless anisotropic shape factor and R is radius of nanoclusters.

In **Figure S13a**, we repeat the same analysis to Yb-Cd systems. A compositional slice of this size-dependent convex hull is taken at the $YbCd_{5.7}$ composition, which is shown in **Figure S13b**. At the $YbCd_{5.7}$ composition, the bulk equilibrium phases are iQC-$YbCd_{5.7}$. Because $Yb_{14}Cd_{51}$ and $Yb_{13}Cd_{76}$ have slightly lower surface energies (35 and 38 meV/Å$^2$) than iQC-$YbCd_{5.7}$, the stability of iQC-$YbCd_{5.7}$ can be size-dependent. However, the intersection of iQC-$YbCd_{5.7}$ and $Yb_{14}Cd_{51}$ + $Yb_{13}Cd_{76}$ occurs at R = 0.17 nm, which is smaller than the unit cell length of $Yb_{14}Cd_{51}$ and $Yb_{13}Cd_{76}$ crystals. Therefore, the size effect is negligibly small while forming iQC-$YbCd_{5.7}$. In **Figure S13c**, we construct the full size-dependent phase diagram of the Yb-Cd system, which is made by projecting the lowest free-energy phase(s) onto the underlying size- and composition-axes.

**Figure S14** illustrates the workflow to generate mixed thermodynamic and kinetic phase diagram of nucleation. Thermodynamic parameters of liquid and solids are from CALPHAD analysis (**Supplementary Information S6**). The $\Delta G_{liquid\text{-}solid}$ is calculated from the gap between the common tangent line of the free energy curve of liquid at given composition and solids at given temperature. The surface energy of solids is calculated from the slope of linear regression of scooped nanoclusters (**Supplementary Information S2**) or the Wulff construction of crystal slabs (**Supplementary Information S5**). The combination of calculated $\Delta G_{liquid\text{-}solid}$ and surface energy derives the nucleation barrier at critical nucleus ($\Delta G_C$) of homogeneous nucleation at given temperature and compositions. The free energy of at critical nucleus ($\Delta G_C$) of homogeneous nucleation was calculated from the following equation.[34]

$$\Delta G_C = \frac{4(\gamma\eta)^3}{27\left(\frac{\rho \Delta G_{liquid-solid}}{N_A}\right)^2}$$

where $N_A$ is the Avogadro number. The $\Delta G_{liquid\text{-}solid}$ of the alloy systems was obtained from the free energy difference between tangent line of liquid at the targeted fraction and solids at the alloy composition. The repetition of the calculation with different temperatures and compositions reveals which solid phase has the lowest nucleation barrier at given temperature and composition. Coloring these phases onto the thermodynamic phase diagram is the mixed thermodynamic and kinetic phase diagram of nucleation.

We perform a CALPHAD assessment of the Yb-Cd liquid free-energy, $G_{liquid}$. We apply the common tangent construction on $G_{liquid}$ to calculate $\mu_{Yb,liquid}$ and $\mu_{Cd,liquid}$, from which we calculate $\Delta G_{liquid-solid}$ to possible intermetallic phases in the thermodynamic temperature-composition phase diagram (**Figure S15a**) except $\beta YbCd_2$ and $Yb_3Cd_8$, of which crystalline structure is not defined. When combined with the surface energies of each phase, **Figure S15b** shows the relative nucleation barriers of each solid phase from a liquid composition of iQC-$YbCd_{5.7}$, as a function of undercooling from the melt. iQC-$YbCd_{5.7}$ has the lowest nucleation barrier at T > 558.88 K.

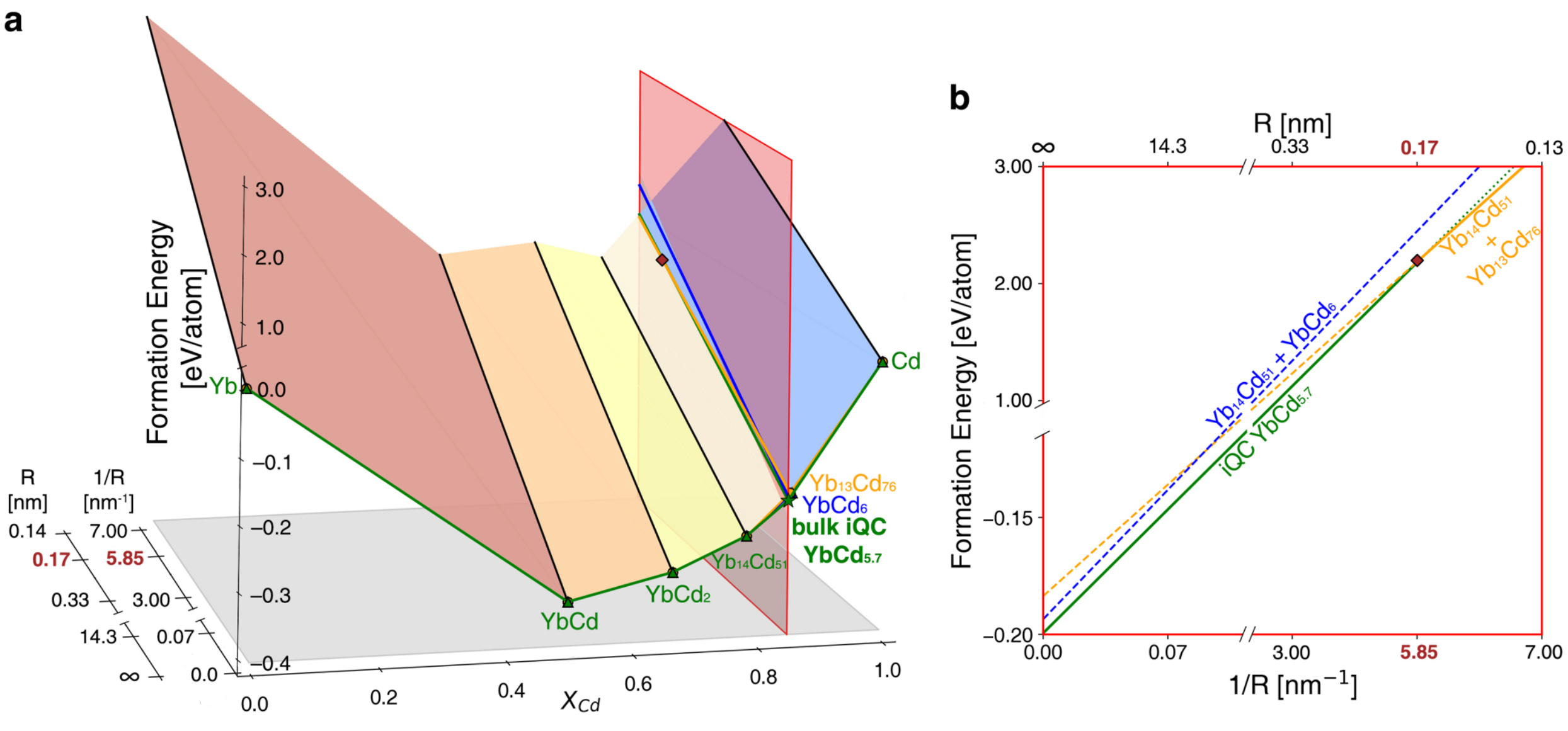

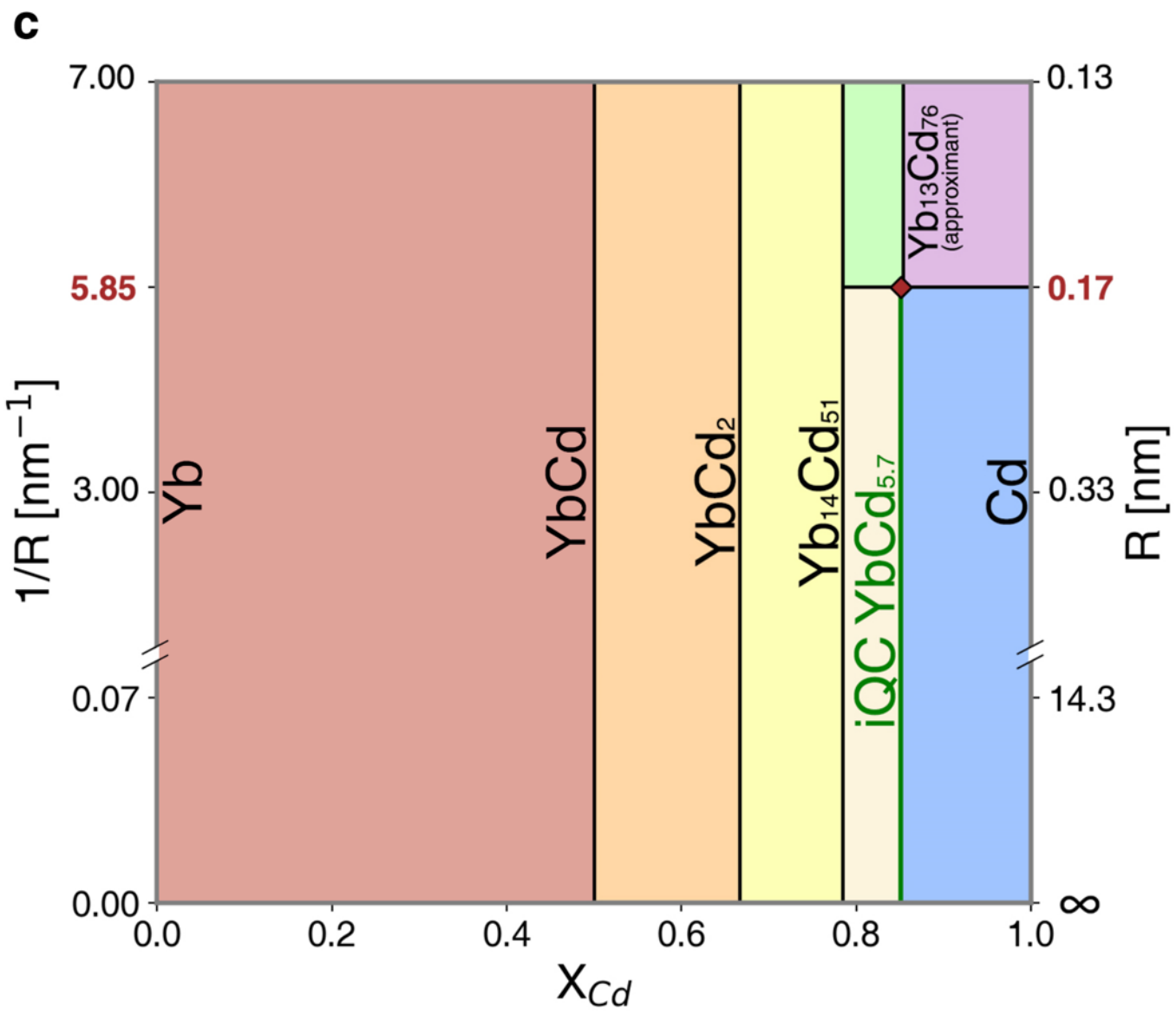

**Figure S13.** **a,** Expansion of 2D convex hull to 3D hull of Yb-Cd system. The green line is the connection between the biggest nanosphere and bulk iQC-$YbCd_{5.7}$ from **Figure 1c**. The orange line on the plane is the line at the composition of intersection point marked brown diamond. **b,** Energy diagram as a function of 1/R at the intersection point. The bold straight lines represent the phase(s) with the lowest energy for a given 1/R. **c,** Size-dependent phase diagram with various Cd compositions. Colors filled between vertical lines represent the phase coexistence region of two end members.

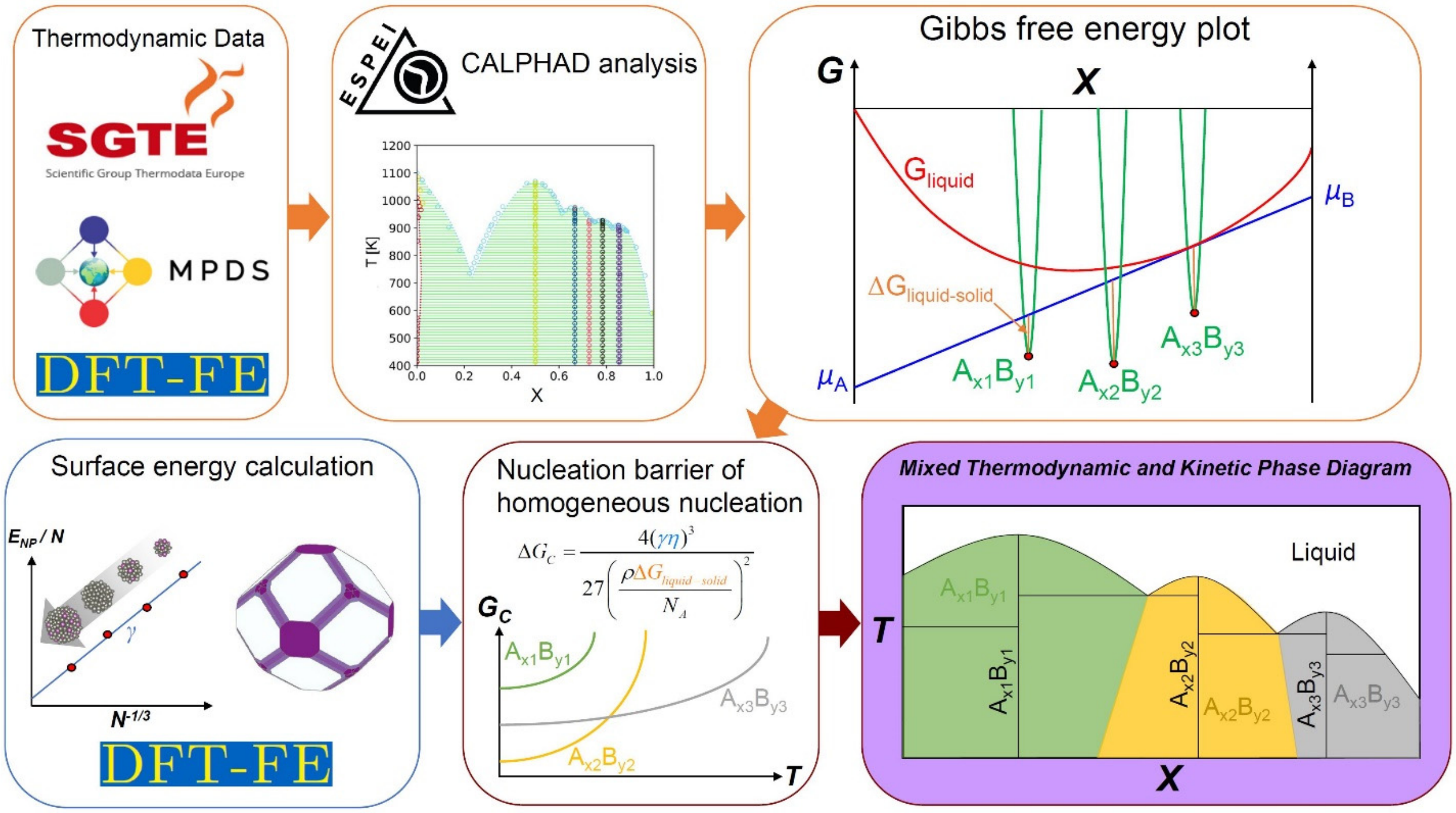

**Figure S14.** Schematic illustration of the workflow to generate mixed thermodynamic and kinetic phase diagram of nucleation.

In **Figure S15c**, we construct a mixed thermodynamic and kinetic phase diagram for the Yb-Cd system. iQC-$YbCd_{5.7}$ is nucleation-preferenced over a wide range of temperatures and near $YbCd_{5.7}$ liquid compositions. Experimental synthesis of iQC-$YbCd_{5.7}$ and $YbCd_6$ was near the congruent and peritectic point according to the thermodynamic temperature-composition phase diagram.[18,26] Our kinetic phase diagram shows the lowest nucleation barrier for iQC-$YbCd_{5.7}$ at its congruent melting point, as well as at lower temperatures within the $YbCd_6$ + liquid peritectic region. The other Yb-Cd solid phases have the lowest nucleation barrier near their solid compositions. From these results, the kinetic selectivity of phases in the Yb-Cd system can be anticipated from the thermodynamic phase diagram, unlike in the Sc-Zn system.

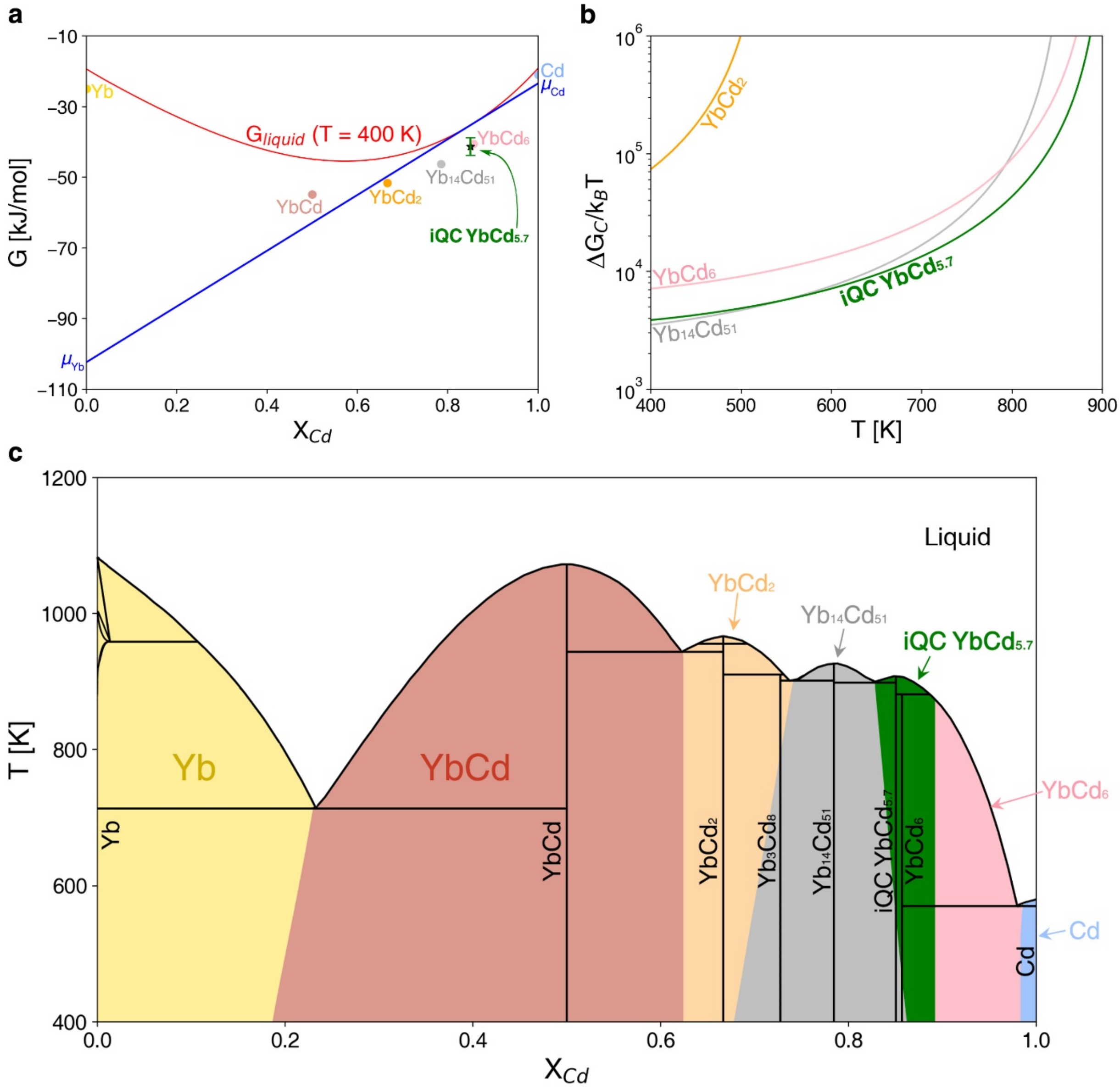

**Figure S15.** **a,** Illustration of common tangent construction in Yb-Cd system. The tangent line of $G_{liquid}$ is drawn at the iQC-$YbCd_{5.7}$ composition and T = 400 K. **b,** Calculated free energy barrier of the critical nucleus at iQC-$YbCd_{5.7}$ composition with different temperature. **c,** Kinetic phase diagram of nucleation on Yb-Cd system. The black lines and compositions represent the equilibrium phase diagram and colored regions below the liquid indicate the areas where the nucleation barrier for a certain phase is the lowest.